**This draft manuscript is distributed solely for purposes of scientific peer review. Its content is deliberative and predecisional, so it must not be disclosed or released by reviewers. Because the manuscript has not yet been approved for publication by the U.S. Geological Survey (USGS), it does not represent any official USGS finding or policy.**

**Title:**

Comparing NASA Discovery and New Frontiers Class Mission Concepts for the Io Volcano Observer (IVO)

**Short Title:**

Spacecraft Mission Concepts for Io

**Authors:**

Christopher W. Hamilton[1,2], Alfred S. McEwen[2], Laszlo Keszthelyi[3], Lynn M. Carter[2], Ashley G. Davies[4], Katherine de Kleer[5], Kandis Lea Jessup[6], Xianzhe Jia[7], James T. Keane[4], Kathleen Mandt[8], Francis Nimmo[9], Chris Paranicas[10], Ryan S. Park[4], Jason E. Perry[2], Anne Pommier[11], Jani Radebaugh[12], Sarah S. Sutton[2], Audrey Vorburger[13], Peter Wurz[13], Cauê Borlina[14,15], Amanda F. Haapala[10], Daniella N. DellaGiustina[2], Brett W. Denevi[10], Sarah M. Hörst[14], Sascha Kempf[16], Krishan K. Khurana[17], Justin J. Likar[10], Adam Masters[18], Olivier Mousis[19], Anjani T. Polit[2], Aditya Bhushan[6], Michael Bland[3], Isamu Matsuyama[2], John Spencer[6]

**Affiliations:**

[1] Corresponding author: chamilton@arizona.edu

[2] Lunar and Planetary Laboratory, University of Arizona, 1629 E. University Blvd. Tucson, AZ 85721, USA

[3] U. S. Geological Survey, Astrogeology Science Center, 2255 N. Gemini Dr., Flagstaff, AZ 86001, USA

[4] Jet Propulsion Laboratory, California Institute of Technology, 4800 Oak Grove Drive, Pasadena, CA 91109, USA

[5] Division of Geological and Planetary Sciences, California Institute of Technology, Pasadena, CA 91125, USA

[6] Southwest Research Institute, 1301 Walnut Street, Suite 400, Boulder, CO 80302 USA
[7] Department of Climate and Space Sciences and Engineering, University of Michigan, Ann Arbor, MI 48109, USA
[8] Planetary Systems Laboratory, NASA Goddard Space Flight Center, Greenbelt, MD 20771 USA
[9] Department of Earth and Planetary Sciences, University of California – Santa Cruz, Santa Cruz, CA 95064, USA
[10] Johns Hopkins University, Applied Physics Laboratory, 11100 Johns Hopkins Road, Laurel, MD 20723, USA
[11] Earth and Planets Laboratory, Carnegie Science, Washington, DC 20015, USA
[12] Department of Geological Sciences, Brigham Young University, Provo, UT 84602, USA
[13] Space Science and Planetology, Physics Institute, University of Bern, 3012 Bern, Switzerland
[14] Department of Earth and Planetary Sciences, Johns Hopkins University, Baltimore, MD 21218 USA
[15] Department of Earth, Atmospheric, and Planetary Sciences, Purdue University, West Lafayette, IN 47907 USA
[16] Laboratory for Atmospheric and Space Physics, University of Colorado Boulder, CO 80303 Boulder, USA
[17] Department of Earth, Planetary and Space Sciences, University of California at Los Angeles, Los Angeles, CA 91356, USA
[18] The Blackett Laboratory, Imperial College London, Prince Consort Road, London SW7 2AZ United Kingdom
[19] Centre National de la Recherche Scientifique (CNRS), Laboratoire d'Astrophysique de Marseille (LAM) LAM, Aix-Marseille Université, Marseille, France

**Abstract**

Jupiter's moon Io is a highly compelling target for future exploration that offers critical insight into tidal dissipation processes and the geology of high heat flux worlds, including primitive planetary bodies, such as the early Earth, that are shaped by enhanced rates of volcanism. Io is also important for understanding the development of volcanogenic atmospheres and mass-exchange within the Jupiter System. However, fundamental questions remain about the state of Io's interior, surface, and atmosphere, as well as its role in the evolution of the Galilean satellites. The *Io Volcano Observer* (*IVO*) would address these questions by achieving the following three key goals: (A) Determine how and where tidal heat is generated inside Io; (B) Understand how tidal heat is transported to the surface of Io; and (C) Understand how Io is evolving. *IVO* was selected for Phase A study through the NASA Discovery program in 2020 and, in anticipation of a New Frontiers 5 opportunity, an enhanced *IVO-NF* mission concept was advanced that would increase the Baseline mission from 10 flybys to 20, with an improved radiation design; employ a Ka-band communications to double *IVO*'s total data downlink; add a wide angle camera for color and stereo mapping; add a dust mass spectrometer; and lower the altitude of later flybys to enable new science. This study compares and contrasts the mission architecture, instrument suite, and science objectives for Discovery (*IVO*) and New Frontiers (*IVO-NF*) missions to Io—and advocates for continued prioritization of Io as an exploration target for New Frontiers.

## 1. INTRODUCTION: IO AS A PRIORITY FOR PLANETARY EXPLORATION

Io's tidal heating drives hundreds of intensely active volcanic systems, which contribute to a global average heat flow that is over twenty times larger than the Earth's (Matson et al. 1981; McEwen et al. 2004; Veeder et al. 2012; Davies et al. 2024a). This makes Io a critically important exploration target for addressing fundamental questions related to early planet formation and interior evolution, as well as for investigating feedbacks between evolving orbital resonances, tidal heating, extreme volcanism, volcanogenic atmospheres, and the broader history of the Jovian system (de Kleer et al. 2019a; Keane et al. 2021a). For these reasons, and more, Io has been consistently identified as a high priority science target within all three National Academies of Sciences Decadal Surveys for Planetary Sciences (National Research Council 2003, 2011, 2022).

This study considers scientific priorities for Io exploration that could be addressed through near-term mission opportunities within NASA's Discovery and New Frontiers (NF) programs. Missions to Io, within either of these mission classes, would result in outstanding and impactful measurements that have the potential to transform our scientific understanding of Io, and tidal heating processes more generally. However, at each level, there are trade-offs between science goals, mission complexities, and costs. This study facilitates comparisons between potential Discovery- and NF-class missions to Io by identifying scientific priorities, capabilities and requirements to obtain measurements that would fundamentally enhance our understanding of Io.

NASA's Discovery mission program is open to all planetary destinations in the Solar System, and missions to Io have been proposed at least four times. In 2020, the *Io Volcano Observer* (*IVO*) was selected for a Phase A study. *IVO* was judged as selectable, but was not selected for a new start. Io is also one of the allowed targets for New Frontiers 5 (NF-5), but the NF-5 call has not been opened at the time of writing. An enhanced version of the *IVO* mission concept *(IVO-NF*) was ready to be proposed for NF-5 in 2023, but the Announcement of Opportunity (AO) has been delayed by several years, as has the next Discovery AO. In 2024, the Committee for Astrobiology and Planetary Sciences (CAPS) is reconsidering which of the NF-5 and NF-6 destinations to recommend to NASA for the delayed NF-5. The Principal Investigator (PI)-managed cost cap for New Frontiers is typically double that of Discovery.

This paper describes and compares the science that could be achieved by *IVO* (Discovery-class) and *IVO-NF* (New Frontiers-class) missions. These mission concepts are designed to orbit Jupiter and make multiple close flybys of Io. Keane et al. (2021b) describe a range of other mission concepts for Io, including an emphasis on small satellites. McEwen et al. (2023) also describe a potential Io orbiter, and Ogliore et al. (2023) describe an Io sample return concept. Additionally, Thomas et al. (2021) described an *IVO*-like concept with greater emphasis on mass loss and energy flows. From the Jet Propulsion Laboratory (JPL) Planetary Science Summer School, several *IVO*-like mission concepts have been presented (e.g., Suer et al. 2017; Hanley et al. 2024). The variety of proposed mission concepts for Io demonstrates the community's broad interest in a dedicated mission to Io, and we consider *IVO* and *IVO-NF* to offer the best combination of affordability and science return, in their respective mission classes, to address priority science questions outlined in multiple Planetary Science Decadal Surveys.

## 2. PRIORITIES AND PRIOR OBSERVATIONS

### *2.1. Io is a High Priority for Solar System Exploration and Planetary Sciences Research*

Io is an intensely active volcanic world (Figure 1, Figure 2) with a global average heat flux of at least 2 $Wm^{-2}$ (Lainey et al. 2009, and prior references). Io's high heat flux is tied to numerous oddities concerning the interior, surface, and atmospheric characteristics of Io and its broader relationship to other bodies within the Jupiter System. For instance, high eruption rates led to rapid rates of resurfacing, crustal subsidence and recycling, as well as the existence and modification of an atmosphere that in turn impacts the nearby space environment. Io also has important interactions with Europa and Ganymede through tidal resonance action, which affects the shape of these bodies, their interiors, and surface geology.

Io has been widely recognized as an especially important object for future study in past Planetary Science Decadal Surveys, with dedicated Io missions included among the New Frontiers mission concepts in 2003 and 2011 (National Research Council 2003, 2011). Measurements enabled by repeated close Io flybys would address many cross-cutting science themes from these documents. For example, from the "Vision and Voyages for Planetary Science in the Decade 2013–2022" Survey (National Research Council 2011), an Io mission would address "Workings of Solar Systems" by providing new data on the evolution of volatiles in the outer Solar System as well as on the internal chemical and physical processes that shape planets and moons. It would also support studies that use the Jovian system as a model for exoplanet systems. An Io flyby mission was also described as critical to understanding the Laplace resonance that drives tidal heating not just on Io, but on Europa and Ganymede as well. This has significant implications for the "Planetary Habitability" theme as we seek to better understand the characteristics and evolution of subsurface oceans on these worlds. Material lost from Io, including oxidants and sulfur species (Thomas 2022), may help to sustain the habitability of the Galilean ocean worlds by providing key nutrients (Becker et al. 2024). Finally, an Io mission would address the "Building New Worlds" theme by improving our understanding of early planet evolution. Io's high temperature eruptions and possible heat-pipe internal structure may be an analog for the early (Hadean) Earth and other terrestrial planets including some exoplanets (Moore et al. 2017).

In the most recent 2023 Decadal Survey, "Oceans, Worlds, and Life" (OWL; National Research Council 2022), Io again features prominently as a unique and key planetary object for understanding the evolution of planets and planetary habitability. In Chapter 5, "Solid Body Interiors and Structures", Io is described as important for understanding magma reservoirs, melt generation, and possibly magma oceans, on tidally heated rocky bodies (Question 5.1). As with "Visions and Voyages" (National Research Council 2011), determining Io's mode of heat transfer is viewed as a key question because of the relevance to heat-pipe planets including the early Earth. As the most volcanically active object in the Solar System, it is important to identify, map and classify volcanic and tectonic features on Io's surface to place constraints on the styles of volcanism, rates and cycles of activity, origins of mountains, and lithospheric properties (Question 5.3, Question 5.6). Io's continuous volcanic activity also impacts its atmospheric composition, and the $SO_2$ exchange processes are currently not well understood (Question 6.4). Io is also extremely important for understanding how circumplanetary systems form and evolve, including their habitability. Chapter 8 of OWL describes the need to obtain a better understanding of interior structure and tidal response of the Jupiter moons including Io, as well as how tidal systems have evolved (Question 8.1). The mountains on Io are again cited as enigmatic features that are critical to understanding planetary tectonics, as is the need to understand the heat transport mode and relationship of the eruptive processes to tides (Question 8.2). Finally, outgassing from Io contributes to the plasma and dust environment at Jupiter and affects the dynamics of the magnetosphere, but the sources, sinks and mass transport over time are still not well understood and need additional measurements (Question 8.4).

### *2.2. Previous Observations*

NASA's *Pioneer 10* and *Pioneer 11* made the first flybys of Jupiter in 1973 and 1974, but it was not until *Voyager 1* and *Voyager 2* passed through the Jovian System in 1979 that humanity achieved its first close view of Io. These images revealed widespread sulfur deposits on Io's surface (Sagan et al. 1979) and the first evidence of active volcanism beyond Earth (Morabito et al. 1979). These observations confirmed Peale et al. (1979)'s earlier prediction that tidal dissipation would drive silicate melting and volcanism on Io, but Io also includes towering mountains (Carr et al. 1979) that are incompatible with initial models that predicted a thin lithosphere (Peale et al. 1979). Perhaps equally surprising was what was not seen on Io—impact craters. The paucity of impact craters requires that the surface of Io be resurfaced at a globally averaged rate of ~1 cm/year (Johnson et al. 1979). This rapid resurfacing rate leads to a unique method of expelling heat from the interior of Io—heat-pipe tectonics (O'Reilly & Davies 1981; Moore 2001). Tidal heating within Io goes into melting rock, the molten rock is transported to the surface via volcanism, and the heat is released as the lava freezes. Rock is brought back into the interior through burial and subsidence. While Io is the only body in our Solar System that currently undergoes strict heat-pipe tectonics, it is plausible that most rocky planets were in this regime early in their history (Moore et al. 2017).

The next close-up views of Io came from the *Galileo* mission as it orbited Jupiter from 1995 to 2003. Most of what is known today about Io's volcanism originated from the *Galileo* mission dataset. While the activity at individual volcanoes had changed since 1979, many volcanic centers imaged by Galileo remained close to the same locations (McEwen et al. 1998) and hundreds of new active volcanoes were identified. *Galileo* observations acquired at visible wavelengths using the Solid State Imaging (SSI) experiment (Belton et al. 1992) as well as at infrared wavelengths using the Near Infrared Mapping Spectrometer (NIMS; Carlson et al. 1992; Lopes-Gautier et al. 2000) and Photopolarimeter-Radiometer (PPR; Russell et al. 1992) confirmed ground-based telescopic observations (Johnson et al. 1988) that suggested much of the Io's volcanism involved silicates, not sulfur, with mafic to ultramafic compositions indicated (Davies et al. 1997, 2001; McEwen et al. 1998; Davies &

Veeder 2023). Styles of volcanism observed on Io were similar to basaltic eruptions on Earth, but typically much grander in scale (Davies 2007). From hundreds of observations acquired by *Galileo*'s imaging instruments in the visible and infrared, *Galileo* data produced detailed analyses of the type, magnitude, distribution, and morphology of Io's volcanic activity over much of Io's surface, as well an inventories of plumes, surface changes, mountains, and paterae. By 2015, 250 thermal sources had been identified from all available data (Veeder et al. 2015).

The global view built up from *Voyager*, *Galileo*, and telescopic observations suggested that Io's short-lived, very-high flux eruptions happen more frequently at high latitudes, while long-lived, steady-state eruptions appeared to be more common near the equator (McEwen & Soderblom 1981; Cantrall et al. 2018). This, and other evidence, was most consistent with tidal heating being more prominent in the upper mantle but with some non-trivial heating deeper within Io (Tackley et al. 2001; Hamilton et al. 2013; Tyler et al. 2015; Matsuyama et al. 2022). Furthermore, volcanic centers concentrated at the sub- and anti-jovian hemispheres, revealing the direct importance of tidal heating on Io's volcanism (Radebaugh et al. 2001; Schenk et al. 2001). The volume of material uplifted in Io's mountains required the lithosphere to be at least 12 km thick on average, but a value of 20–50 km was most likely (Jaeger et al. 2003).

At the end of the *Galileo* era, analysis of the spatial distribution of volcanic centers and mountains showed that: (a) volcanic centers exhibit a strong preference for formation adjacent to mountains (Schenk et al. 2001; Radebaugh et al. 2001; Jaeger et al. 2003), but on a global scale, there may be subtle variations in the concentrations of mountains and volcanoes that are spatially anti-correlated (Schenk et al. 2001; Kirchoff et al. 2011); (b) the distribution of volcanoes on Io is more uniform than random (Hamilton et al. 2013), which implies that self-organization processes affect magma transport through the lithosphere and volcano spacing at the surface; and (c) volcano clusters (Hamilton et al. 2013) and heat flux distributions (Veeder et al. 2012; Davies et al. 2015) generally agreed with heat flux patterns predicted by asthenospheric-dominated tidal heating models (Tackley et al. 2001; Beuthe 2013), but volcano distributions and heat flow patterns are shifted 30–60° to the East. This offset can be explained by the presence of a magma ocean within Io (Tyler et al. 2015), but additional measurements are needed to uniquely constrain four possible models for Io’s interior (Figure 3). Two of these models involve solid-body tidal dissipation, which would either concentrate heating and silicate melting in the asthenosphere or deep-mantle. The other two models involve fluid-body tides, which would occur within the asthenosphere and may involve either a continuous magma ocean or a magmatic sponge composed of interconnected partial melt that can move through permeable lithic matrix. Figure 3 summarizes the key observables that would be expected for each of these four scenarios, and examined by *IVO* or *IVO-NF*. Figure 4, Figure 5, and Figure 6 provide additional detail regarding the measurements that are needed to quantitatively distinguish between these alternatives.

Geophysical observations by the *Galileo* magnetometer also point to the presence of a global magma ocean within Io (Keszthelyi et al. 1999; Khurana et al. 2011). However, there are questions raised about these interpretations (Moore 2001; Blöcker et al. 2018). In terms of Io’s internal structure and composition, *Galileo* data for Io is limited, but the data place useful constraints on the size of Io’s core, which is most likely completely liquid (Breuer et al. 2022). The difference between the measured Moment of Inertia (MoI) for Io and the value of a sphere with a homogeneous density (MoI = 0.4), implies that Io has a core that is denser than its mantle, and Moore et al. (2007) estimate that Io’s core forms 0.37 to 0.52 of its mean radius of 1821.3 km—assuming shell densities in the range of silicate rocks and core densities corresponding to a range for iron–iron sulfide mixtures. This corresponds to a possible core thickness of approximately 675 to 950 km. However, significant uncertainties remain because the amount of sulfur that is sequestered in the core is not known (Schubert et al. 2004). In fact, processes that move sulfur and sulfur compounds in, on, and around Io all remain largely unknown and understanding Io’s sulfur cycle is an important priority for future investigations.

Thermal emission observations can also be used to infer eruption temperatures and magma compositions (Davies et al. 2011). Partial melts erupted from Io’s mantle are likely either mafic (i.e., basaltic) or ultramafic (i.e., komatiitic), but reliable measurements of their eruption temperature are only possible in specific and relatively demanding circumstances. For instance, lava fountains can be detected by their thermal emission (e.g., by *Galileo*; Davies et al. 2001) and even Earth-based telescopes (Veeder et al. 1994; Marchis et al. 2002; de Kleer et al. 2014; de Pater et al. 2014), but their characteristics exhibit high temporal variability, both thermally and spatially. Terrestrial analog studies show that to determine eruption temperature to within 100 K, multispectral imaging of lava fountains requires ≤100 m/pixel resolution and <0.1 s between different color observations (Davies et al. 2011). In contrast, lava tube skylights, likely common on Io but not yet identified in existing data, are very stable both temporally and spatially (Davies et al. 2016, 2017). Lava tube skylights may therefore provide valuable targets for future high spatial resolution measurements of lava temperature on Io.

The *Cassini* spacecraft flew past Jupiter on the way to the Saturn system and, over the course of three days in early 2000, data were acquired using the Imaging Science Subsystem (ISS) with measurements at ultraviolet (UV) to near-infrared (Near-IR) wavelengths for global eclipse imaging. Several prominent hotspots were observed, including Loki Patera (13°N, 309°W), Pillan (12°S, 243°W) and Pele (19°S, 255°W; Radebaugh et al. 2004; Allen et al. 2013). Color images of the eclipse showed that the auroral emissions come from a mix of species (Geissler et al. 2004). Compositional data of nanograins ejected by Io's volcanoes and entrained in the solar wind were obtained by the Cosmic Dust Analyzer (CDA). Sodium chloride (NaCl) was identified as the major particle constituent, accompanied by sulfurous as well as potassium bearing components (Postberg et al. 2010).

The *New Horizons* spacecraft flew by the Jupiter system in 2007, passing Io at a distance of 2.3 million kilometers (Spencer et al. 2007). It obtained 17 images of Io using the Multicolor Visible Imaging Camera (MVIC), which is a four-color (visible to NIR) camera, 190 images using its high-resolution panchromatic camera, the Long-Range Reconnaissance Imager (LORRI), and 9 spectral cubes from the Linear Etalon Infrared Spectral Array (LEISA). A total of 54 thermal emission sources were identified in LORRI and MVIC data, including the first short timescale observations of 7 individual ionian volcanoes, which were imaged on time scales ranging from seconds to minutes (Rathbun et al. 2014). 37 Ionian volcanoes were observed using LEISA, which is sensitive to infrared wavelengths between 1.25 and 2.5 µm (Tsang et al. 2014). The best observations of active volcanoes were taken for Tvashtar (63°N, 124°W), Xihe (58°S, 293°W), and East Girru (22°N, 245°W). The volcanoes generally exhibited stable brightnesses on these short timescales, but East Girru did exhibit variation on the order of minutes to days, increasing by 25% in just over an hour and decreasing by a factor of 4 over 6 days. A cluster of hotspots around Io’s sub-jovian point—that was also observed by the *Galileo* Solid State Imager (SSI)—was observed in LORRI data while Io was eclipsed by Jupiter; however, most of these were not detected by LEISA in the near-infrared. These are interpreted as gas emissions from interactions with the magnetosphere, however the exact mechanism remains unclear (Rathbun et al. 2014).

NASA’s *Juno* spacecraft is now orbiting Jupiter in a high inclination orbit that has presented opportunities to image Io at visible and IR wavelengths from a polar perspective. These observations have completed the global survey of active volcanic centers down to a spatial resolution of <10 km/pixel (e.g., Mura et al. 2020; Davies et al. 2024a,b). *Juno* infrared data were combined with previous analyses to identify 343 discrete hot spots (Davies et al. 2024b). Using volcanic heat flow as a proxy for tidal heating-induced global heat flow, the volcanic thermal emission distribution and hot spot distribution correlate poorly with deep mantle radially-integrated heating models, but perform better with asthenospheric tidal heating, and slightly better than the latter with the heat flow from a global magma ocean, although the correlations for these “shallow” models are generally weak (Davies et al. 2024b). However, *Juno* cannot map the lower temperature anomalies (≲200 K) and background heat flow that account for much of Io’s heat loss. A pair of close flybys (1,500 km) in late 2023 and early 2024 provided an opportunity for *Juno*’s JunoCam visible wavelength imager (Hansen et al. 2017) to image areas of Io poorly covered by previous missions (Williams et al. 2024), providing vital geological context to the measurements of high-temperature thermal emission from Io’s polar region volcanoes.

Near-IR telescopes, including those equipped with adaptive optics, continue to provide regular coverage of Io’s volcanic activity based on near-infrared emission (e.g., de Kleer et al. 2019b; Tate et al. 2023); such campaigns provide the most regular coverage for detection of rare outburst events. Ground-based optical observations as well as ultraviolet data from the *Hubble Space Telescope* (*HST*) provide key information on Io's atmosphere and ionosphere for understanding mass loss processes (e.g., Roth et al. 2014, Schmidt et al. 2023). The next generation of ground-based near-infrared telescopes (e.g., the *Extremely Large Telescope* [*ELT*], *Thirty Meter Telescope* [*TMT*], *Giant Magellan Telescope* [*GMT*]) will provide dramatic improvements to spatial resolution (~30 km resolution at Io, at λ = 1 μm), as well as to the sensitivity to spectral lines due to the larger collecting area. The *James Webb Space Telescope* (*JWST*) has also observed Io across the near- and mid-infrared; although *JWST* saturates on Io over some key wavelength ranges (≳13 μm and parts of the near-IR), valuable data can still be obtained in the unsaturated spectral regions. While observations are at a relatively low spatial resolution (~350 km/pixel in the near-IR and ~750 km/pixel in the mid-IR), infrared spectra have been obtained from 0.7 to 5.3 μm at high spectral resolution (λ/Δλ = ~3000), obtaining spectra of gas emission from Io's most powerful volcanoes (de Pater et al. 2023). The *Atacama Large (sub-)Millimeter Array* (*ALMA*) can map molecular species in Io's atmosphere with high sensitivity to trace the composition of Io's bulk atmosphere as well as in plumes (de Pater et al. 2020; Redwing et al. 2022; de Kleer et al. 2024); planned upgrades to ALMA in the upcoming decades will increase its ability to map many species simultaneously, and may improve its sensitivity, which would enable detection of new species.

Recently, a new imaging system, SHARK-VIS (Conrad et al. 2024) has been used to observe Io. SHARK-VIS is a three-band visible-wavelength (400–900 nm) imager which, utilizing the *Large Binocular Telescope* (*LBT*]) and adaptive optics, obtained images of Io in late 2023 and early 2024 at 80 km/pixel. Observations at such spatial resolution were previously only possible from spacecraft. SHARK-VIS provides vital context for IR observations.

NASA's *Europa Clipper* mission (Pappalardo et al. 2024), and the European Space Agency (ESA)'s *Jupiter Icy Moons Explorer* (*JUICE*) mission (Grasset et al. 2013)—will be the next missions to the Jupiter System. They will be able to observe Io from near or outside Europa's orbit, but from mostly equatorial perspectives. These spacecraft will have opportunities to provide vital coverage for extending the timeline of activity at Io’s volcanoes and for monitoring changes on Io’s surface in the 2030–2035 timeframe. However, it is essential to fly close to Io to make key measurements of tidal gravity, magnetic induction, and atmospheric and plume composition, and to acquire higher-resolution remote sensing data.

## 3. THE DISCOVERY-LEVEL *IO VOLCANO OBSERVER* (*IVO*) MISSION CONCEPT

Distant monitoring by telescopes and spacecraft has taught us much about Io’s active volcanism and magnetospheric interactions (de Kleer et al. 2019b). However, some of the most fundamental questions that remain unanswered require getting close to Io. These issues include: (1) understanding the geophysics via gravity and magnetometer measurements; (2) obtaining high-resolution remote sensing over a range of wavelengths to observe the body shape, rotational motions, landscape and active eruptions; (3) measuring Io’s orbital evolution; and (4) directly measuring the composition of the atmosphere and plumes. Furthermore, a dedicated mission is needed to ensure that orbits are designed to achieve key objectives such as measuring tidal gravity and libration and with science instruments designed for the intense radiation environment and associated noise at Io. These issues would all be addressed by a future *Io Volcano Observer* (*IVO*) mission.

### *3.1. Science Goals and Objectives of IVO*

Future missions to Io should address the priorities outlined in the Planetary Science Decadal Surveys, which can be simplified to three key goals (Table 1): Goal A—Determine how and where tidal heat is generated inside Io; Goal B—Understand how tidal heat is transported to the surface of Io; and Goal C—Understand how Io is evolving.

A Discovery-class *IVO* mission concept can address all three of these goals in the Baseline (but not Threshold) mission, whereas the *IVO-NF* mission concept (Section 4) can accomplish these goals in the Threshold mission, with substantial “bonus” science. Note that only the Threshold requirements are necessary to achieve the minimum science acceptable for the investment (i.e., what is required for NASA to continue the mission), whereas the Baseline requirements include all of the science and performance requirements necessary to achieve the full science objectives of the mission. A NF-class mission could also support twice as many close flybys, including some lower-altitude flybys, and accommodate new or improved science instruments. Given the complexity of Io, the extra and unique data from NF could prove essential to answering key questions and making new discoveries.

### *3.1.1. Goal A: The goal of determining how and where tidal heat is generated inside Io*

Goal A ("Determine how and where tidal heat is generated inside Io") would constrain the thermo-rheological state of Io's mantle (Figure 3 and Figure 4), which is a key unknown in limiting tidal heating and magma ascent models. The goal would be addressed by acquiring improved measurements of: (Objective A1a) tidal $k_2$, (Objective A1b) libration amplitude, (Objective A1c) magnetic induction, and (Objective A1d) lava compositions and their thermophysical properties (Figure 7). Tidal response places a strong constraint on a planetary body's interior, especially as described by the parameter $k_2$ (Munk & MacDonald 1960). The tidal response of an Io with a shallow magma ocean ($k_2 > 0.5$) is much larger than for an Io without ($k_2 = \sim 0.1$; Bierson & Nimmo 2016). Since $k_2$ represents a change in Io's gravity field, it can be measured by a series of flybys, as was done for Titan with *Cassini* (Durante et al. 2019). Libration amplitude provides an independent test for whether or not Io has a detached lithosphere over a magma ocean. As with Enceladus (Thomas et al. 2016), repeat imaging of the same point on a surface at different epochs can reveal a libration with an amplitude of hundreds of meters, smaller than Io's likely libration (Van Hoolst et al. 2020). Combining measurements of $k_2$ and libration allow both the lithospheric thickness and the rigidity to be determined (Van Hoolst et al. 2020). Time variations in Jupiter's magnetic field drive electrical currents and create induced magnetic fields that can provide information about Io's interior by constraining the properties and geometry of the inductive layer, which may correspond to a global subsurface magma ocean (Khurana et al. 2011). The fourth independent method of probing the state of Io's interior is to measure the composition and temperature of the lavas and exsolved gasses. These four independent ways of inferring information about Io's interior combine to provide a powerful framework for understanding tidal dissipation processes within Io, which is critical for understanding its interior evolution and volcanic history.

*3.1.2. Goal B: Understand how tidal heat is transported to the surface of Io*

The key objectives of Goal B ("Understand how tidal heat is transported to the surface") are to (Objective B1) determine Io's lithospheric structure, and (Objective B2) determine where and how Io is losing heat. These objectives are important because a planetary body's lithospheric structure is a consequence of how the body loses its heat. Objective B1 can be addressed through establishing better constraints on (Objective B1a) lithospheric thickness and rigidity. If a magma ocean is present, the $k_2$ tidal response of Io depends on the thickness and rigidity of the lithosphere (Wahr et al. 2006). For libration, the tradeoff is different. A thicker shell with higher rigidity has the same libration amplitude as a thinner shell with lower rigidity (Van Hoolst et al. 2013). Thus, the measurements from Objectives A1a and A1b can tightly bound the allowable combinations of lithospheric thickness and rigidity (Figure 3 and Figure 4). An additional approach involves (Objective B1b) topographic mapping to provide regional and local tests of the volcanic heat-pipe model (Moore 2001) and the relationships between faults, mountains, and lithostatic stresses (Ahern et al. 2017), which would relate to patterns of heat flux and subsidence. *Galileo*-era topography does not support the presence of a magma ocean (Gyalay & Nimmo 2024) but the strength of this conclusion is limited by the small amount of data available.

The second objective (B2) is to determine where and how Io is losing heat. The global distribution of heat flow is a key constraint on where tidal heat is generated within Io, especially if there is no magma ocean. Tidal heating models predict that higher heat flow at the poles of Io indicate heating deep in the mantle, while heat loss focused at certain longitudes near the equator indicate relatively shallow heating (Figure 3) (Segatz et al. 1988; Tackley et al. 2001; Hamilton et al. 2013; Tyler et al. 2015; Steinke et al. 2020). Two key ways to address this objective include: (Objective B2a) generate improved endogenic heat flow maps with better resolution and coverage and sensitivity, particular in the polar regions of Io; and measure (Objective B2b) active eruption parameters because the style, vigor, and heat flux from volcanic eruptions are closely tied to its magma temperature, composition and processes of ascent. Understanding where and how heat is transported to the surface reveals the inner workings of Io's geologic engine, and understanding the sulfur cycle is critical to interpretation of fractionation of sulfur isotopes (de Kleer et al. 2024). Of particular interest is how the volcanic, tectonic, and sulfurous cycles operate and interact in Io's lithosphere. Basic parameters that need to be measured are: (1) the thickness and rigidity of the lithosphere; and (2) the mechanisms and spatial distribution of heat flow.

Lithospheric properties for Io are best constrained by a combination of different geophysical and geodetic techniques. In particular, the combination of measurements of libration, gravity responses to tides, and the induced magnetic field would tightly constrain lithospheric thickness and rigidity if Io has a detached lithosphere over a magma ocean. These methods are complementary, with very different axes of degeneracy across parameter space (Figure 4). This means that, while each individual technique has significant uncertainties, the results of a multi-method study leads to very tight constraints. However, one of the key assumptions in the data inversion is that of globally uniform lithospheric properties. By making detailed topographic measurements around tall mountains the flexural response of the lithosphere on a more local scale can be obtained. A distributed set of flexural response measurements would provide a critical test of the assumption that the lithosphere is similar across much of Io.

Previous missions (see Section 2.2) show that Io's volcanic heat flow is prodigious and have provided rough maps of how it is distributed across Io (Veeder et al. 2012, 2015; Davies et al. 2015). These maps have since been refined by the inclusion of *Juno* data (Davies et al. 2024b). Analysis of these data show, while the number of hot spots per unit area is uniform between polar regions (>60° latitude) and lower latitudes, high-temperature volcanic thermal emission from polar regions is only half that from lower latitudes, and such thermal emission from the north polar region is twice that of the south polar region (Davies et al. 2024b). However, the total heat flow including lower temperatures remains unknown. The nature of this apparent dichotomy is as yet unexplained, and might be a function of lithospheric structure. Refining these measurements further would be useful, but a dedicated Io mission would aim much higher. A particularly vexing problem is that the observed volcanic heat flow is ~55% (Davies et al. 2024b) of Io's modeled total heat flow (105 ± 12 TW; Veeder et al. 1994). It is possible that significant heat is transported by other mechanisms. Widely distributed cryovolcanism involving sulfur and sulfurous compounds, which *Galileo* was insensitive to, may transport a significant amount of heat. Heat flow from near-surface intrusions and buried volcanic deposits would be hard to detect at local scales as surface temperature variations might be small and subtle. Nevertheless, Io's unaccounted for thermal emission (in terms of surface distribution) is ~47 TW, the equivalent of Earth's entire endogenic heat flow (Davies & Davies, 2010). Furthermore, a crucial test of the heat-pipe model would be to place upper limits on the heat flow via conduction. This requires separating the effects of insolation from endogenic heat flow, which can be done by observing surface temperatures at multiple times of day along with a contemporaneous high-spatial resolution Bond albedo map and robust determination of local surface thermal inertia.

### *3.1.3. Goal C: Understand how Io is evolving*

Io is a dynamic world and a key to understanding why it is so dynamic involves understanding to what extent Io is in a steady state, or part of a more rapidly evolving system. Goal C ("Understand how Io is evolving") would be addressed via two critical objectives: (Objective C1) measure Io's orbital parameters to constrain its (Objective C1a) rate of orbital change; and (Objective C2) determine the current rate of volatile loss from Io. Objective C2 involves four components: (Objective C2a) measure the composition and abundances of neutral species in Io's atmosphere; (Objective C2b) monitor volcanic plumes; (Objective C2c) monitor emissions in eclipse; and (Objective C2d) measure the variability of the plasma and magnetic fields in the Jupiter system. Measuring isotope ratios also has great promise towards understanding Io's evolution (de Kleer et al. 2024) if the fractionation processes are well understood (Hughes et al. 2024). Measurements from both gas and dust mass spectrometers are key to understanding fractionation processes.

Io's atmosphere provides an ideal laboratory for studying the processes most relevant to the evolution of volatiles because production and loss of atmospheric species connects directly with primordial conditions (e.g., Pollack & Witteborn [1980] for Io and Mandt et al. [2014, 2017] for Titan and Pluto, respectively). Determining Io's current volatile loss rates requires using models validated by measurements of the composition, abundance, and distribution of neutrals in Io's atmosphere and exosphere as a function of altitude and time of day with the Ion and Neutral Mass Spectrometer (INMS). Io's atmosphere is currently known to be composed of sulfur- and oxygen-bearing species (e.g., $SO_2$, SO, $S_2$, $O_2$, S, and O (Pearl et al. 1979; Lellouch et al. 2007; Spencer et al. 2000; McGrath et al. 2000; Geissler et al. 2004) as well as salts (e.g., NaCl, KCl, Na, K, and Cl; Lellouch et al. 2003; Moullet et al. 2015).

Although not yet detected in Io's atmosphere, the most common form of sulfur is $S_8$ and studies of Io's atmospheric composition should not exclude this species. High resolution imaging of Io in eclipse is particularly valuable because it enables detection of emissions from Io's atmosphere and plumes to determine their spatial and temporal variability. Measurement of Io's plasma environment would also offer a key constraint of mass flux from Io. The interaction between magnetospheric plasma and Io's atmosphere results in ~1000 kg/s of volatiles being lost from Io (e.g., Broadfoot et al. 1979; Hill et al. 1979; Dessler 1980; Delamere & Bagenal, 2003; Hikida et al. 2020; Bagenal & Dols, 2020), forming a neutral torus (e.g., Burger & Johnson, 2004; Koga et al. 2018; Smith et al. 2022), and via ionization by magnetospheric electrons a plasma torus (e.g., Lellouch et al. 2003; Yoshikawa et al. 2014; Koga et al. 2019), where some of these ions can be transferred to Europa (Becker et al. 2024). When combined, these observations described above will inform our understanding of the processes illustrated in Figure 8.

INMS neutral composition measurements, covering a wide range of altitudes, local times, latitude, longitude, and plasma interactions along with emission observations during eclipse, will validate our three-dimensional model calculations of atmospheric loss rates based on simulations of ion and neutral profiles as a function of altitude, latitude, and longitude (Wurz et al. 2021). This multi-atmosphere model is successfully validated for Mercury (Wurz et al. 2019; Gamborino et al. 2019), Venus (Gruchola et al. 2019), and the icy Galilean moons Europa, Ganymede, and Callisto (Vorburger et al. 2015, 2019, 2022, 2024; Vorburger & Wurz 2018) and has recently been adapted for Io for the *IVO* Concept Study (Vorburger et al. 2021). A review of the processes releasing particles into the atmosphere and their loss is given by Wurz et al. (2022). The *IVO* mission would improve constraints on models of volcanic output rates through monitoring of volcanic plume activity on every flyby along with the condition of the magnetosphere throughout the *IVO* orbit. Furthermore, observations of emission during eclipse will help to constrain condensation rates at night.

### *3.2. IVO Instrument/Measurement Summary*

*IVO*'s Baseline mission experiments (Table 2) are designed to address the three goals described in Section 3.1 and can be accommodated as a Discovery-class mission. All instruments have significant heritage from other missions to high radiation-environments, such as *Europa Clipper*; *JUICE*; *MErcury Surface, Space ENvironment, GEochemistry, and Ranging* (*MESSENGER*); and BepiColombo.

Per the Discovery AO, a Technology Demonstration Option (TDO) Reflective UV Spatial Heterodyne Spectrometer (RUSHeS) was also proposed to complement the baseline instruments on *IVO*. The primary objective of RUSHeS was to demonstrate the improved capabilities of spatial heterodyne spectrometers in measuring emission line signatures at both high etendue and high spectral resolving power in flight. RUSHeS contributed to the scientific goals of *IVO* Discovery by measuring the neutral-ion emission lines in the atmospheres of the Galilean moons, the Jovian aurora, and the interplanetary medium.

The Student Wide Angle Camera (SWAC) was also included as an optional Student Collaboration experiment in the Discovery *IVO* proposal. SWAC was proposed as a 50° × 25° Field of View (FOV) panchromatic (450–1000 nm) framing camera with a 46.2 mm focal length and an F/6 refractor, similar to the Origins, Spectral Interpretation, Resource Identification, and Security – Regolith Explorer/Apophis Explorer (*OSIRIS-REx/APEX*) Camera Suite (OCAMS): MapCam and SamCam (Rizk et al. 2018). The SWAC detector would have been the same as the SRI International 4k × 2k Complementary Metal-Oxide-Semiconductor (CMOS) detector used by *Europa Clipper*, and planned for *IVO*'s Narrow Angle Camera (NAC). The primary objective of SWAC was to provide experiential education to university students at upper division undergraduate and graduate levels. Students were to be involved with the camera design, build, testing and calibration through coursework, internships, and employment. Students would play key roles in the SWAC build, test, and delivery to Assembly, Test and Launch Operations (ATLO). At Io, students would have contributed to the planning and scientific analysis of SWAC images. The scientific objective of SWAC was to offer complimentary capabilities to the NAC. Planned capabilities of the SWAC included imaging Io's limb, overexposed to show stars, as a check on NAC measurements to determine the libration amplitude to distinguish between a coupled and decoupled surface; acquiring Io limb profiles and stereo coverage at identical illumination conditions to accomplish topographic mapping of km-scale features; providing additional surface coverage to identify active eruption parameters; and monitoring plume activity by observing Io's illuminated limb and terminator.

Data products for the Discovery *IVO* mission were carefully planned to achieve all science objectives described in Section 3.1. The data management plan was based on extensive experience at the University of Arizona developing and operating the ground data systems for *Mars Reconnaissance Orbiter* (*MRO*)/High Resolution Imaging Science Experiment (HiRISE), *Trace Gas Orbiter* (*TGO*)/Colour and Stereo Surface Imaging System (CaSSIS), and *OSIRIS-REx/APEX*. Data acquired during operations would be downlinked from the spacecraft via the Deep Space Network (DSN) to the Mission Operations Center (MOC) and then to the *IVO* Operations Center (IVOC). Automated pipelines would process the raw data through radiometric calibration. These pipeline-processed data would be released to the Planetary Data System (PDS) every three months. Derived products (Table 3) would be produced by team members with the relevant expertise. The proposed data management plan emphasized the development of a Planetary Spatial Data Infrastructure (PSDI) for Io including foundational products to increase the discoverability and interoperability of *IVO* data products (Laura et al. 2017; Bland et al. 2021; Williams et al. 2021). In addition to the expected return from the baseline instruments at Io, data products from the technology demonstration opportunity, student collaboration instruments (e.g., SWAC), and asteroid flyby enroute to the Jovian system were included in the plan should they have been selected.

## 4. THE NEW FRONTIERS 5 (NF-5)-LEVEL *IO VOLCANO OBSERVER* (*IVO-NF*) MISSION CONCEPT

The higher PI-managed cost cap of the NASA New Frontiers (NF) program enables new science that is difficult to accomplish in Discovery. In 2023, we completed an Applied Physics Laboratory (APL) Concurrent Engineering (ACE) study and concluded that the following set of enhancements was well within the draft 2023 PI-managed NF cost caps ($900M for development phases and $300M for post-launch phases):

(1) Nominal mission with 20 Io encounters rather than 10, using an improved radiation design.
(2) Use of Ka-band telecom system for more rapid data downlink, so the 20 flybys can occur with the same 3.5-year tour as the 10-flyby Discovery mission. The total Io data volume return is roughly doubled.
(3) Addition of a baseline Wide Angle Camera (WAC) similar to that developed for *Europa Clipper* (Turtle et al. in press) for color and stereo mapping.
(4) Addition of a dust mass spectrometer like the Surface Dust Analyzer (SUDA) on *Europa Clipper* (Kempf et al. in press).
(5) Lowered altitude of some later flybys to enable new science, pending safety analysis with SUDA and other data to understand dust hazards.

The 3.5-year Jupiter tour duration for both *IVO* and *IVO-NF* is driven by gravity science requirements to enable four close encounters at the ideal locations over Io near both periapse and apoapse. This is needed for an unambiguous measurement of tidal $k_2$. The fuel needed to support the NF tour is actually less than in the 10-encounter Discovery mission because each orbit is smaller and experiences less solar perturbation. Furthermore, the cadence of operations is more efficient, without long time gaps between some encounters, thereby substantially improving the science return per dollar.

### *4.1. Science Goals for IVO-NF*

#### *4.1.1. Static gravity science*

Io’s interior structure remains enigmatic. One of the most powerful approaches for probing the interior structure of planetary bodies is by measuring their gravity fields, usually through Doppler tracking of spacecraft as they flyby or orbit their target. *Galileo* flybys provided only hints of the gravity field of Io—limited to measurements of Io’s bulk density and longest-wavelength gravity terms (see Keane et al. 2022 for a review). The *Juno* flybys are anticipated to provide the first constraints on Io’s time-dependent gravity, but due to the large flyby distances (>1,500 km altitude), it is unlikely that *Juno* will be able to meaningfully constrain Io’s gravity field at higher spatial resolution (beyond spherical harmonic degree $n = 2$). Measuring Io’s static (i.e., time-independent) gravity field at higher resolution is important as it opens up new investigations of Io’s subsurface structure. For example, the combination of static gravity and topography can enable inferences about spatial variations in crustal thickness and the compensation state of the lithosphere. There is extensive literature on this topic for a range of planetary bodies, including for the Moon (Wieczorek et al. 2013), Mars (Goossens et al. 2017), Ceres (Ermakov et al. 2017), Enceladus (Hemingway et al. 2018; Park et al. 2024). For Io, different models for heat production and melt transport make differing predictions for how Io’s crustal thickness may vary spatially (e.g., Spencer et al. 2020).

To evaluate how well *IVO-NF* would recover Io’s static gravity field, we performed a dynamical simulation of the gravity science experiment. In general, the quality of the gravity field recovery depends on the number of flybys (more is better), flyby altitude (lower is better), and observing geometry (prefers variation). *IVO-NF*’s Doppler data are highly sensitive to the change in the spacecraft velocity in the line-of-sight direction from a DSN station, thereby allowing recovery of the gravity signal affecting the spacecraft motion and other non-gravitational forces. Based on realistic assumptions, detailed simulations are performed to assess the expected recovery of accuracy of Io’s static gravity field. In this analysis, the dynamical model and estimated parameters were nearly identical to a previous study for Europa (Park et al. 2011, 2015), except using estimated gravity parameters for Io. Specifically, for the baseline, the estimated parameters included three position parameters ($\sigma_a$ = 100 km), three velocity parameters ($\sigma_a$ = 1 m/s), three constant acceleration parameters ($\sigma_a = 5 \times 10^{-11}$ km/s$^2$), Io's GM (where GM is is the Gravitational Constant [G = $6.67 \times 10^{-11}$ m$^3$ kg$^{-1}$ s$^{-2}$] times the mass of Io [~$8.93 \times 10^{22}$ kg], which is ~$5.96 \times 10^3$ km$^3$/s$^{-2}$; $\sigma_a$ = 10% of current estimate), $k_2$ ($\sigma_a$ wide open), and 20 × 20 normalized spherical harmonics coefficients for the static gravity field (Kaula law for $\sigma_a$), where $\sigma_a$ represents the a priori uncertainty of an estimated parameter. The position, velocity, and acceleration parameters were estimated for each flyby. The three constant acceleration parameters represent the accelerations in the radial, transverse, and normal directions and their nominal values were set to zero. These accelerations were used to model the errors in the non-gravitational forces, such as solar and planetary radiation pressures, spacecraft thermal emission, etc. Since flybys occur over a relatively short time frame, three constant accelerations should be a good representation of the errors in the non-gravitational models. For the a priori uncertainty of the 20 × 20 gravity field, the Kaula power law (Kaula 2000) of $19 \times 10^{-4}/n^2$ was considered (where $n$ is the spherical harmonic degree), which is about ten times larger than the gravity coefficients expected from Io's observed long-wavelength topography (White et al. 2014) with no compensation, making our assumption conservative. If the topography of Io is compensated, this a priori Kaula law would be lowered by a factor of ~10–50, which would improve the recovery of $k_2$. For the degree-2 coefficients, the a priori uncertainties were taken from the *Galileo* results (Jacobson 2013), but inflated by a factor of 30 to make our simulation robust.

Figure 9 shows the expected quality for a global static gravity field from 20 flybys for *IVO-NF*. The range of plausible Io gravity fields are shown as a red/blue swath, where the magnitude of the expected gravity field is higher if Io's topography is uncompensated, and lower if compensated. *IVO-NF*'s capability is shown as the thick black line. On average, a degree-5 gravity field can be recovered (1,150 km half-wavelength resolution). For comparison, this is approximately the same resolution as the current gravity fields for Titan (Durante et al. 2019) and Ganymede (Gomez Casajus et al. 2022). Importantly, a degree-5 gravity field of Io would be sufficient to determine if observed long-wavelength patterns in Io’s volcanism, tectonism, and heatflow—which all have strong signals at degree-2 to degree-5 (e.g., Kirchoff et al. 2011; Keane et al. 2022)—are related to subsurface structure. While a global degree-5 gravity field would be insufficient to directly resolve individual volcanoes or mountains, it is important to note that *IVO-NF*’s flybys would enable higher quality determinations of local gravity field in select regions where the *IVO-NF* flyby groundtracks overlapped. It may be possible to improve the quality of the gravity field in these targeted regions by a factor of two or more.

*4.1.2. Isotopes and Io’s long-term evolution*

The state of Io’s interior today must be the result of extensive processing because of the moon’s intense tidal heating, which drives high rates of interior melting, eruptive activity, outgassing, and recycling of crustal material back into the mantle via burial and subsidence. Nonetheless, Io’s history volcanism remains poorly understood, including whether the degree of tidal heating has changed over Io’s history and how its mantle and magma composition have evolved over time. Furthermore, it is unclear whether Io formed as dry as it is now (Mousis et al. 2023) or formed wet and lost its water later as a result of tidal heating or other processes (Canup & Ward 2002; Sasaki et al. 2010;  Ronnet et al. 2017; Bierson & Nimmo 2020; Bierson et al. 2023). Formation models suggest that the Galilean satellites formed from the accretion of solids, likely in pebble form, within the Jovian circumplanetary disk, which had a lifespan of at most a few million years after Jupiter's formation (Canup & Ward 2009; Ronnet et al. 2017; Mousis et al. 2023). These solids could have originated from the protosolar nebula or possibly condensed in situ within the Jovian circumplanetary disk (Mousis & Gautier 2004; Prinn & Fegley 1981). The current consensus is that the Galilean moons underwent Type I or II migration within the circumplanetary disk, which was halted when they were captured into orbital resonance. (Peale & Lee 2002; Alibert et al. 2005; Sasaki et al. 2010; Batygin & Morbidelli 2020; Madeira et al. 2021), implying that Io has been tidally heated for Gyrs. However, Io’s rapid resurfacing erases evidence of anything except the past Myr of its history (Johnson et al. 1979). The abundances of the stable isotopes of the volatile-forming elements have the potential to provide information on Io’s long-term history, similarly to how the D/H ratio has been used to infer the mass-loss history of Mars and Venus (Donahue et al. 1982; Webster et al. 2013). *IVO-NF* would greatly enhance isotope science compared to *IVO* because of more and lower-altitude encounters and the addition of a dust mass spectrometer like SUDA (Kempf et al. 2024).

$SO_2$ is the primary constituent of Io's atmosphere and coats Io's surface in frost form. Recent work using ground-based submillimeter observations has shown that Io's sulfur inventory has a larger $^{34}S/^{32}S$ ratio than other Solar System bodies, which was used to infer that Io may have been volcanically active and losing mass for billions of years (de Kleer et al. 2024). The measurement of other stable isotopes can test and extend this type of interpretation, but only the more abundant isotopes are detectable to Earth due to the sensitivity of the available instruments. Gas and dust mass spectrometer instruments have the potential to measure the suite of isotopologues of $SO_2$ as well as other atmospheric species that have been detected in lesser abundances (SO, $S_2$). In particular, the four stable isotopes of S ($^{32}S$, $^{34}S$, $^{33}S$, $^{36}S$, in order of relative abundance) and the three stable isotopes of O ($^{16}O$, $^{18}O$, $^{17}O$) can constrain the amount of mass Io has lost over time, test the fractionation mechanisms proposed by de Kleer et al. (2024) and Hughes et al. (2024), and perhaps determine whether Io's mass loss has always been primarily in the form of $SO_2$ or whether other molecules such as $H_2O$ have also been lost in significant quantities. Mass degeneracy between the isotopologues is an important consideration (e.g., $^{34}SO_2$ vs. $^{32}S^{16}O^{18}O$), with modeling required to measure the full set of isotopologues.

Measuring the triple oxygen isotope system would additionally enable placing Io within the larger context of Solar System bodies for which triple oxygen isotopes have been measured (e.g., Ireland et al. 2020), which has the potential to inform our understanding of where within the protoplanetary disk Io's volatiles predominantly condensed. Although H has not yet been detected at Io in any form, a measurement of D/H at Io would add another constraint on the origin of Io's volatiles. If Io's measured D/H ratio were significantly heavier than known Solar System reservoirs, this could provide strong evidence that Io formed wet and has lost significant water over its history. In contrast, if Io's measured D/H ratio is close to the protosolar value, it would imply that its building blocks condensed in an initially warm and dense circumplanetary disk (Mousis 2004; Horner et al. 2008).

*4.1.3. Magnetic field generation in Io, crustal magnetism, and electrical conductivity*

The existence of an internal dynamo in Io remains undetermined, with a reported maximum surface magnetic field of <50 nT (Khurana et al. 1998; Kivelson et al. 2004). While this value is low when compared to other moons and planets in the Solar System (e.g., Ganymede's surface field ranges from ~750 to ~1,500 nT; Kivelson et al. 2004), the upper limit of the magnetic field generated at the surface could help differentiate between different mechanisms and locations for the dynamo. For present-day Io, it is unlikely that a magnetic field could be generated through a thermal dynamo in its magma ocean or in the core because tidal heating hinders the extraction of heat from the core (Wienbruch & Spohn 1995). However, a mechanism for magnetic field generation that could be actively present in Io is a mechanically driven dynamo resulting from tidal forcing (Landeau et al. 2022). This mechanism has not been extensively investigated, however past studies have suggested that moons with ellipticities $>1 \times 10^{-7}$, such as Io (Oberst & Schuster 2004; Gao & Huang 2014), could produce vigorous dynamos. Due to its large tides, Io is the ideal place to test the hypothesis. Depending on the magnetic properties and the geometry of the conductive layers measured by *IVO-NF*, different locations and mechanisms for the dynamo might generate different fields that could match the <50 nT surface field previously reported. Overall, *IVO-NF* would provide a unique opportunity to investigate this mechanism to drive dynamos that could have implications for other moons and planets in the Solar System.

Jupiter’s magnetic field traps plasma over a wide energy range. The main species of the magnetospheric plasma are electrons, protons, and various charge states of oxygen and sulfur. As the magnetospheric plasma overtakes Io in its orbit, previously bound neutrals can become liberated. Bagenal & Dols (2020) estimate the overall neutral loss rate as 250–3000 kg/s and provide a complete discussion of the key processes, including how these neutrals can become ionized. While Io is a major source of oxygen and sulfur, it is also believed to supply minor species to the surrounding medium. Observed and postulated examples include Na, K, and Mg. Detection of these species will give us clues about Io and its interior.

Electrical conductivity can be used to describe the present-day thermal state of Io’s interior (e.g., de Kleer et al. 2019a) and inform about its cooling. To model the conductivity of Io’s interior, electrical measurements in the laboratory on Io analogs at relevant conditions are needed. Although the exact composition and structure of Io remains to be determined, assumptions can be made based on previous studies. We estimate the electrical conductivity of Io at depth (Figure 10) using the thermal and compositional profiles of Breuer et al. (2022). Io’s crustal conductivity ranges from $10^{-3}$ to 16 S/m over 300–1500 K, assuming electrical studies of basalts at 1 atm (solid and molten state; Rai & Manghnani 1977) and S-free and S-bearing andesitic melts at high temperature (Waff & Weill 1975; Pommier et al. 2023, respectively). The high conductivity values in the lower crust reflect the presence of silicate melt at high temperature, which is consistent with upwelling magma. The melt content in the mantle was varied from 0 to ~30% (Breuer et al. 2022 and references therein), and its conductivity taken at 1500 K. The electrical dataset on S-bearing natural melts is limited to one study, with experiments conducted at 2 GPa on highly reduced andesitic glasses and melts containing up to 5 wt.% S (Pommier et al. 2023). The conductivity of reduced S-bearing melts varies from 0.7 to 2.2 S/m, with a small increase in conductivity by a factor of ~3 as the sulfide content increases from 0 to 5 wt.%. These conductivity values are in the range of S-free natural silicate melt conductivities. We note that the redox state of Io remains to be constrained, and it is possible that melts in Io are oxidized, with a slightly different dependence of conductivity to composition.

The solid part of Io’s mantle consists of highly sheared olivine, as a result of tidal deformation. Depending on the shear strain and the direction of measurement (parallel to the main shear direction or orthogonal to the shear plane), olivine conductivity was found to vary by a factor of 8 (Pommier et al. 2018). The bulk conductivity ($\sigma_{bulk}$) of a partially molten mantle can be calculated using the geometric mean, which considers a random distribution of the phases, with arbitrary shaped and oriented volumes (Glover 2015):

$$\sigma_{bulk} = \sigma_{melt}^{X_{melt}} \times \sigma_{olivine}^{(1-X_{melt})}$$ [Eq. 1]

with $\sigma_{melt}$ the conductivity of S-bearing silicate melt, $\sigma_{olivine}$ the conductivity of sheared olivine, and $X_{melt}$ the volume fraction of melt. We assume that the effect of pressure on conductivity is negligible across Io's thin mantle. At 1500 K, increasing the amount of melt in the asthenosphere from 0 to 30 % increases bulk conductivity by a factor of ~10. A sharp increase in conductivity is observed at the core-mantle boundary. Considering a core composition in the Fe-S(-Ni) system, conductivity is expected to range from ~$10^5$ to $2.5 \times 10^5$ S/m, using data at 2 GPa from Saxena et al. (2021). The relatively insulating mantle compared to the conductive core might affect core cooling, and future modeling work is needed to assess the effect of tidal heating on the conductivity-depth profile and the thermal evolution of Io.

### *4.2. Additional Science Instruments for IVO-NF*

*IVO-NF* adds a Wide Angle Camera (WAC) as a Baseline instrument. Although it was desired to have the *Europa Clipper*'s Europa Imaging System (EIS) WAC on Discovery *IVO*, it would have exceeded nominal power requirements, and thus was proposed as the student collaboration option (Section 3.2). New Frontiers-class power requirements could accommodate the EIS WAC, allowing it to be incorporated as a baseline instrument. Heritage was maintained by utilizing the EIS WAC design and spare detectors, but with filters optimized for science at Io. In addition to a clear filter for panchromatic imaging, seven color filters were selected to complement the NAC color and provide continuity with *Galileo* (Belton et al. 1992) and *Voyager* (Smith et al. 1977) color data. Similar to EIS WAC, the detector layout allowed for three-line pushbroom stereo, clear framing, and pushbroom color imaging with Time Delay Integration (TDI). Multi-band ratios could be used to constrain lava temperatures. Products from the NAC and WAC instruments would be comparable to those planned for EIS NAC and WAC (Table 4).

Following evaluation of several additional science instruments, the top priority that emerged was a dust mass spectrometer similar to SUDA on *Europa Clipper* (Kempf et al. 2024). This instrument would provide several major advances. First, by better understanding the dust environment around Io, we can have greater confidence about flying closer to Io's surface, which has major science advantages as described in Section 4.1.2. Nominally we would fly as close as 50 km, but certainly closer than the 200 km minimum of *IVO* in the Discovery mission. Second, by measuring dust composition, we would obtain additional constraints on surface compositions, including whether or not silicate lavas have ultramafic compositions. Third, the instrument measures the composition of the nanograins that are entrained in the volcanic plumes (e.g. Postberg et al. 2024). Fourth, this instrument can measure isotopes from a different reservoir than the gas measured by INMS, providing a key test of models for fractionation processes (e.g., Hughes et al. 2024).

Several other additions to *IVO-NF* were considered. Small spacecraft could fly into active vents, returning video and mass spectra (Wurz et al. 2017). However, the technology readiness level for this type of experiment is too low for a Baseline experiment, so further consideration was deferred until a Phase A study, perhaps as a TDO. Other instruments considered included UV spectrometer, near-IR spectrometer, advanced pointing camera (Park et al. 2020), energetic particle detector, passive radar experiment (Steinbrugge et al. 2021), and other experiments. All of these instruments would return valuable data on an *IVO-NF* mission, but maintaining large margins in mass, power, and cost were favored in the initial studies.

### *4.3. The IVO-NF Tour*

The *IVO-NF* journey to the Outer Solar System would transit through the asteroid belt, providing an opportunity to observe objects in the outer Themis family—a reservoir of icy, "active" asteroids not yet visited by spacecraft (Landis et al. 2022). These asteroids, some of which display visible tails and comae, may provide a link between asteroids, comets, and ocean worlds (Lauretta et al. 2024). A flyby of a Themis-family asteroid by *IVO-NF* would advance our understanding of these primitive bodies using a New Frontiers-class payload. Likely remnants of an icy planetesimal's catastrophic disruption, the Themis family objects are rich in water ice and organic compounds, making them ideal targets for addressing key questions in planetary science and astrobiology (Landis et al. 2022). During an asteroid flyby, the *IVO-NF* Thermal Mapper (TMAP), NAC, SWAC, INMS, and SUDA instruments could obtain measurements, including isotopic measurements and fluxes of outgassing materials, to characterize silicate composition and assess surface features indicative of aqueous alteration and volatile activity. Detailed observations would include high-resolution imaging, spectroscopy, and thermal measurements, enabling the determination of the asteroid's crater-retention age, surface composition, and thermal inertia. The expected data return will provide unprecedented in situ measurements of icy asteroid volatiles, including the first D/H ratios from a Themis-family asteroid, which could help shed light on the origin of Earth's water (Licandro et al. 2012). This flyby will also provide the *IVO-NF* team with a valuable rehearsal to exercise operational tools and processes shortly before (~1 year) the first Io encounter.

The Baseline *IVO* Discovery tour at Jupiter (Figure 13) includes 10 flybys of Io, designed to accommodate a diversity of science campaigns including near-global lit surface coverage, thermal mapping, gravity science and libration measurements, magnetic induction studies, and observations of key sites such as Loki Patera (13°N, 309°W) and Pele (19°S, 255°W). Each flyby is planned to optimize particular science objectives, and the flybys are sequenced to maximize the quality of various measurement types based on system geometries. The overarching priorities of the science tour are largely preserved between the Discovery and *IVO-NF* designs, however, the *IVO-NF* tour is expanded in several key ways.

The Baseline *IVO-NF* tour design is expanded to include 20 Io flybys (Figure 13), enabling more complete illuminated (i.e., daytime) and nighttime / eclipse surface and atmospheric coverage of Io, repeat coverage over a number of hot spots, targeted flybys over additional sites of interest, and enhanced magnetic induction and gravity science campaigns. Reduced minimum altitudes for targeted flybys of Io further enhance science quality for the *IVO-NF* tour. Preserving the tour duration of 3.5 years between the 10- and 20-flyby tours enables an increased density of science acquired, without significantly increasing operations costs. To accommodate the faster flyby cadence (i.e., shorter duration between Io passes), Ka-band is utilized for downlink in the New Frontiers design.

A 20-flyby tour carries commensurate increases in penetrating radiation fluences and, consequently, increases in predicted Total Ionizing Dose (TID) and Total Non-Ionizing Dose (TNID), or Displacement Damage Dose. Initial predicted values (no Radiation Design Margin applied) of 250 krad (Si) at 250 mil Al (6.35 mm Al) and 160 krad (Si) at 500 mil Al (12.70 mm Al) suggest options for shielding and Radiation Hardness Assurance (RHA) approaches. Initial trade studies (considering complexity, mass, cost, etc.) used to determine the value of a radiation vault, similar to those integrated on *Europa Clipper* and *Juno*, against the effectiveness of a prudent layout and localized box-level shielding, indicate that the vault design would concentrate mass, leading to undesirable effects on the spacecraft structure design. Recent years have seen favorable growth in the availability of 100 krad (Si) and 300 krad (Si) Electrical, Electronic, and Electromechanical (EEE) parts, offering confidence in parts selection and effectiveness of upscreening.

Given Io's position deep within Jupiter's radiation belts, the rapid orbital speed at perijove (24–25 km/s for both *IVO* and *IVO-NF*) minimizes dose to the spacecraft per Io encounter. This orbit enables the spacecraft to spend most of its time in a more manageable radiation environment, particularly since the spacecraft speed is slowest at apojove (0.7–1.4 km/s). The large orbital apojove nominally allows for the recovery of some system parts (e.g., through annealing). Particle radiation (>100 keV electrons and ions) has been shown to be very variable close to Io itself. For example, Paranicas et al. (2024) show for the two *Juno* Io flybys, which occurred at almost the same close approach altitude, the radiation can vary by orders of magnitude. In addition, several *Juno* instruments reported a small, but significant decrease in radiation along Io's orbital path and even radially inward and outward of Io's orbit (Paranicas et al. 2024).

In addition to surviving the total dose of the radiation environment, the instantaneous flux of penetrating radiation near Io has to be considered that will cause significant background in the various detectors used in the instruments. For the Neutral gas and Ion Mass spectrometer (NIM) instrument on *JUICE*, the radiation shielding of the detector was optimized for the Europa flybys, and verified at accelerators, to minimize the background from penetrating radiation (Lasi et al. 2017). Similar shielding against penetrating radiation of detectors will be necessary for all detectors to perform measurements with good signal-to-noise (S/N) on *IVO* or *IVO-NF*.

*IVO* and *IVO-NF* science operations are focused primarily within the four days before and after each Io encounter, with the rest of each orbit dedicated to the playback of data from the previous encounter. For most of the encounter period, NAC and TMAP would perform hourly plume and hotspot monitoring observations, as well as three to four, three-hour eclipse sequences. As the spacecraft approaches within one hour of Io, dedicated multispectral scans with TMAP and NAC would be performed, along with clear-filter mosaics to globally map Io, measure libration and acquire volcano movies. Finally, within 10 minutes of closest approach, the spacecraft will be turned so that WAC, NAC, and TMAP may acquire high-resolution pushbroom imaging swaths of Io's surface, with the latter two using the pivot to target individual features. The Dual Fluxgate Magnetometer (DMAG), Plasma Instrument for Magnetic Sounding (PIMS), INMS, and SUDA will also acquire data at their highest rates during the close approach period, collecting key magnetic, plasma, and gas and dust compositional information.

The *IVO* concept tour would fly at altitudes of 300 km or higher except near Pele (200 km), whereas the *IVO-NF* concept would include lower-altitude flybys over active volcanic centers, such Pele Patera (Figure 11), at altitudes as low as 50 km—if deemed safe, and occurring late in the mission after most requirements have been met. *IVO-NF* would fly above active regions with its solar arrays edge-on to minimize potential degradation.

In 2001, *Galileo* flew through a plume at an altitude of 200 km above Thor (39°N, 133°W). The Thor plume had distinct gas and dust components, with the dense optical dust core reaching altitudes of ~100 km. At an altitude of 200 km, the Thor plume was nearly pure gas (Frank et al. 2002). This experience confirms that transit through regions that are nearly pure gas can be done with no significant damage to spacecraft optics or hardware.

The Pele plume deposition pattern includes a large red oval shape extending to ~1000 km in diameter (Figure 11). Models suggest the Pele plume is generated from gas and dust erupting from multiple fissures existing on the lava lake located at the heart of the Pele plume deposition pattern (McDoniel et al. 2015). Pele plume observations indicate that the gas of the plume extends to altitudes of 400–600 km, and *Hubble Space Telescope* and other spacecraft observations of the Pele plume in reflected sunlight have shown dust distributed within the Pele plume at altitudes ~300–400 km from the surface (Jessup et al. 2012). Models designed to replicate the Pele plume surface deposition pattern require a broad range of dust particulate sizes extending from 0.02 to 10 μm in diameter (McDoniel et al. 2015). These models suggest that the smallest dust particulates (i.e., those <0.1 μm) remain collocated with the gas flow for long distances reaching maximum altitudes only ~100 km lower than the expanding gas plume and are later deposited on the surface in the large outer red ring deposits. Dust particles (~0.3–1.5 μm in diameter) are needed to fit the dark deposits interior to the red ring. The vertical altitude and horizontal displacement achieved by these larger dust particles decreases as the particle size increases. *IVO* and *IVO-NF* plan to fly directly over the Pele deposition region traveling in a north-easterly direction from the SW to NE edge of Pele's red ring deposits (Figure 11). Convolution of the navigational tracking of the planned flyover trajectory with the plume structure indicates that the spacecraft will intersect the particulates responsible for Pele's distinct red ring at altitudes of 280–270 km at entry and at ~200 km altitude on exit—spending a total of 9 seconds intersecting the red deposit source region. Our estimate of the dust mass to be encountered by *IVO-NF* during a Pele flythrough is well within the $10^{-14}$ to $10^{-5}$ kg mass encountered by the *Stardust* spacecraft during its Comet 81P/Wild 2 flythrough (Green et al. 2004).

*Galileo* observations indicate that Prometheus (and other plumes like it) are characterized by dense optical dust cores that reach altitudes of ~50–80 km, and have outer gas envelopes that reach altitudes ~2–3× higher than the optical dust core (Geissler and McMillan 2008). *Voyager, Galileo*, and *Juno* images all indicate that the surface deposition pattern at Prometheus includes a bright outer ring of $SO_2$ frost encircling the Prometheus vent region (1°S, 152°W) at a radius of ~200–230 km (Douté et al. 2001), with an inner core of darker material (Strom et al. 1981; Geissler & Goldstein 2004). The Prometheus plume is generated by fresh lava interacting with the $SO_2$ frost covered surface (Kieffer et al. 2000) while a smaller more sulfur-rich plume is exsolved directly from the lava vent (McEwen et al. 2000). The *IVO-NF* trajectory produced in 2023 flies NW directly over the center of the Prometheus plume at an altitude of ~51–60 km, which according to modeling by Zhang et al. (2004), suggests that the spacecraft would only encounter a particle mass of ~$10^{-8}$ to $10^{-6}$ kg, again within the mass encountered by the *Stardust* spacecraft.

A plume has not been detected at Loki (located to the north east of Loki Patera) since the *Voyager* encounters in 1979. The pair of Loki plumes imaged by *Voyager* may have represented a hybrid of Pele and Prometheus plume types (McEwen & Soderblom 1981). The Loki plume deposits appear to be centered on a wandering lava flow, some distance from the lava lake located in Loki Patera (see Figure 1). Red coloration along the edge of Loki Patera suggests surface deposits produced from gas exsolution from the lava lake. Given the proximity of the previously observed Loki plume deposits to Loki Patera, monitoring by *IVO-NF* will be needed for analysis of a potential plume hazard to select the actual flyby altitude for a low pass over Loki Patera, as well as to reconsider altitudes for any other plume flythroughs. Changing the closest approach altitude is a minor perturbation to the trajectory and is easily accommodated. NAC will also establish plume conditions prior to implementing passes above Loki Patera, the Pele lava lake, and the Prometheus lava flows, as well as monitoring the activity and optical characteristics of other active plumes on Io.

### *4.4. Improved Science with the NF-5 Baseline Mission*

With twice as many orbits, improved data transmission rates, and expanded instrumentation (Table 5), *IVO-NF* will improve upon a Discovery-class *IVO* mission concept by addressing important scientific challenges related to: (1) magnetic induction; (2) extreme volcanism; (3) volcanogenic atmospheres; (4) mountains and tectonics; and (5) mass exchange within the Jovian System. Following subsections explore each of these improved science areas in more detail.

*4.4.1. Magnetic induction—Constraining Io's interior melt distribution and conductivity*

Io generates a strong inductive field, potentially explained by an ocean of conductive molten magma (Khurana et al. 2011). The depth to which a signal penetrates a body is characterized by its skin depth $s = \sqrt{\mu_0 \sigma \omega / 2}$, where $\mu_0$ is vacuum permeability, $\sigma$ is the conductivity of the body, and $\omega$ is the signal frequency. The two key frequencies with strong driving signals are the synodic rotation period of Jupiter, $T_{syn}$ (12.95 h, amplitude ~800 nT) and the orbital period of Io, $T_{orb}$ (42.46 h, amplitude ~30 nT). Figure 6 shows the expected response at these frequencies as a function of magma ocean thickness (*y*-axis) and magma ocean conductivity (*x*-axis), which is related to the melt fraction of the magma ocean. Figure 12 shows the error in inverting the magnetic field observations from a 20 flyby mission—based on our iteration 3 trajectories—as a function of the number of flybys included in the inversion. The final error of inversion is ~2.5 nT and individual errors in inverting $T_{syn}$ is 5 nT and inverting $T_{orb}$ is 1 nT. These accuracies will allow us to determine the mean thickness of Io's lithosphere with an accuracy of ~4 km. The conductivity and thickness of the magma ocean would be known to better than 50% of their values. If only 10 flybys were available as in the Discovery class mission, the errors of inversion are roughly a factor of 2 higher and thus all of the magma ocean parameters would be obtained with twice the uncertainties shown here. In these inversions, we assume that plasma moments upstream of Io are available (i.e., density, flow velocity, and temperature) from an accurate plasma instrument such as PIMS, so that the plasma interaction generated magnetic fields can be obtained from magnetohydrodynamic (MHD) simulations and removed with an accuracy of 5 nT or better.

*4.4.2. Extreme volcanism—Implications for Io and understanding the early Earth and other high heat-flux worlds*

Ionian eruptions generally dwarf their terrestrial counterparts (Davies 2007), and the range of volcanic eruption styles observed on Io are indicative of low-silica content magma erupted at high temperatures. Resulting explosive eruptions, lava flows and lava lakes are enormous (Lopes et al., 2004); and provide opportunities to directly observe primordial eruption styles that once helped to shape the early Earth, Moon, and other planetary bodies, but are now extinct due to secular cooling (Matson et al. 1998; Keszthelyi et al. 2006). Additionally, the derivation of lava composition—and the distribution of its composition variation—is of vital importance in understanding Io's interior heating and structure. An important approach to determining silicate composition utilizes the wavelength of the material's Christiansen Feature (Conel 1969). Just as mineralogical maps have been developed for the surface of the Moon using IR measurements around 8 μm, from the *Lunar Reconnaissance Orbiter*'s Diviner Lunar Radiometer Experiment (e.g., Lucy et al. 2021), *IVO*'s Thermal Mapper will be used to infer-silica abundance using five bands in the ~7.0–9.5 μm region to map composition and distinguish between mafic and ultramafic compositions.

The larger number of flybys in *IVO-NF* provides substantially more opportunities to observe volcanic activity at high spatial resolution as well as better temporal monitoring of key sites at moderate spatial resolution. This is invaluable for understanding the variety of eruptive styles operating on Io and will allow more confident interpretation of volcanic activity not seen at high to moderate spatial resolution (e.g., monitoring with Earth-based telescopes). The increased number of close-in observations will also improve the robustness of measurements of the distribution of different types of volcanic activity. This is important for understanding regional variations in volcanic heat flux across Io.

Mapping and understanding Io's heat flow and how this is distributed across Io is key to understanding interior tidal heating (de Kleer et al. 2019a, and references therein). The additional data from the higher number of flybys should allow better constraints to be placed on thermal inertia and the background conductive heat flux through Io's lithosphere. Obtaining thermal data from the same location at different times of day is essential for calculating thermal inertia and the additional flybys significantly improve the robustness of the expected results. Furthermore, the additional observations provide more opportunities to search for and characterize sulfurous volcanism. It is possible that this process plays a non-trivial role in moving heat across the lithosphere.

*4.4.3. Volcanogenic atmosphere—Implications for Io's interior evolution and mass transport within the Jovian System*

Io's atmosphere is predominantly composed of $SO_2$, with a surface pressure on the order of (1–10) × $10^{-4}$ Pa. This atmosphere is fed by Io's rich volcanic activity, although the specific contributions of direct volcanic outgassing versus maintenance through sublimating $SO_2$ frost require further investigation (e.g., Lellouch 2004). The density and temperature of Io's atmosphere exhibit notable variations, influenced by factors such as time of day, latitude, surface frost abundance, and volcanic activity, with fluctuations on the order of ~10–100. In addition, the atmospheric $SO_2$ flux density is strongly diminished during Jupiter's eclipse of Io (Tsang et al. 2016). In contrast to $SO_2$, non-condensable gasses might 'survive' Jupiter eclipse, and buffer, or even prevent, atmospheric collapse, though this still needs to be assessed in more detail. Besides $SO_2$, Io's atmosphere consists of $SO_2$ dissociation products (SO, S, O, and $O_2$; with $O_2$ to be confirmed), and both NaCl and KCl have been observed to be present in Io's atmosphere (e.g., de Pater et al. 2020), though their non-colocation with both SO and $SO_2$ remains a riddle.

Interactions between Jupiter's magnetosphere and the Galilean moons lead to feedbacks that affect the space environment around mass transport within the Jupiter System and into interplanetary space (Bagenal & Dols 2020). For instance, impact of magnetospheric ions on the moons' atmospheres supplies clouds of escaping neutral atoms that populate a substantial fraction of their orbits; and ionization of atoms in the neutral cloud provide the primary source of magnetospheric plasma. Additionally, ion and neutral implantation into Europa's surface ices introduce additional species that can react with endogenic material to produce hydrated alkali sulfates, chlorides, carbonates, complex hydrocarbons, and other products (Thomas 2022). Understanding contributions to the Jovian plasma environment from Io and the other Galilean moons is therefore critical for understanding the broader role of mass transport within the system, with implications for prebiotic molecular assembly and habitability.

*4.4.4. Mountains and Tectonics—understanding resurfacing rates and Io's rock cycle*

Stereophotogrammetric observations via the Baseline mission WAC is a major advantage for tectonic studies for the *IVO-NF* concept. In particular, obtaining detailed topographic models of multiple Ionian mountains, and the surrounding plains, would allow tests of different models for the geometry of the underlying orogenic faults. Some geometries create local extensions around the mountains (Keszthelyi et al. 2022), potentially explaining the observed common juxtaposition of volcanic centers and mountains (Jaeger et al. 2003; Radebaugh et al. 2001). Being able to image faults would be highly valuable, as many are buried under $SO_2$ frosts. Higher-resolution imaging may enable this.

The topography of mountains and the surrounding aprons of debris will also allow investigations of mass wasting processes. Mass wasting must be a key element of Io's rock cycle, complementing volcanism in resurfacing the planet. However, the dearth of existing data has largely limited studies to descriptions of localized processes (Schenk and Bulmer 1998; Ahern et al. 2017). The coverage and spatial resolution of the *IVO-NF* data should allow a comprehensive global inventory of mass wasting processes to be compiled. Recently, a potential small impact crater was identified in high-resolution *Galileo* images (Williams et al. 2023). No other craters have been identified on Io's surface. Accomplishing a more complete, high-resolution crater survey could help yield a better surface and resurfacing age for Io.

*4.4.5. Io's Plasma Environment—understanding mass transport within the Jovian System*

Io ejects ~1000 kg/s of sulfur-rich gasses into the giant magnetosphere of Jupiter, driving million-amp electrical currents that excite strong auroral emissions over the poles of Jupiter (Bagenal et al. 2024; Thomas 2022). The processes involved include atmospheric and surface sputtering, charge-exchange, and photo- and/or electron-impact ionization and subsequent pick-up. Material lost from Io forms a neutral cloud that surrounds and accompanies Io in its orbit about Jupiter. This cloud undergoes electron impact ionization and charge exchange to produce the dense Io plasma torus (up to 4000 electron $cm^{-3}$). The neutral cloud and the Io plasma torus are known to contain S, O, $SO_2$, Na, K, and Cl, but additional species will be discovered by the first mass spectrometer in the system.

How the neutral clouds and plasma vary with Io's volcanic activity are not well understood, but having a dedicated Io mission monitoring both Io and its environment is what is needed. Also, the neutral cloud and plasma torus are large enough to be observed at exoplanets, so understanding the Io–Jupiter System is key to interpretations of exoplanet observations (Oza et al. 2019). Thus understanding Io's plasma environment is important "bonus" science for the mission.

## 5. CONCLUSIONS

Io is a fundamentally important planetary body for understanding tidal heating processes, magma oceans, and the evolution of high heat-flux planets—including the early Earth, volcanic exoplanets, and many other planetary bodies during their early geologic history. Io has additional importance for understanding the evolution of the Jovian System, including the assembly of the Laplace resonance between Io, Europa, and Ganymede, and mass transport between the Galilean satellites and into interplanetary space. A dedicated mission to Io would obtain unique information to address critically important community goals regarding the origin and evolution of Io, as well as its role within a broader evolution of the Jovian System. The key difference between *IVO* at the Discovery-level and *IVO-NF* at the New Frontiers-level is that *IVO* would "Follow the Heat", whereas *IVO-NF* would address this goal and go a step further to also "Follow the Mass" by developing a comprehensive view of mass transport from Io's interior to its surface, atmosphere, and throughout the Jovian System—including affecting the surface environment and potential habitability of Europa.

Io is a geologically active and exciting world, and its exploration is fundamentally important for addressing many high priority science questions for the coming decades of space exploration. Both *IVO* (Discovery-class) and *IVO-NF* (New Frontiers-class) would provide outstanding science return and transform our understanding of Io. However, *IVO-NF* would more than double the science return of a Discovery-class mission by completing at least twice the number of flyby, including two new science instruments, return twice the total data volume, and have opportunities to conduct lower altitude flybys of active volcanic systems to acquire unique higher-resolution datasets and obtain direct information about the composition of Io’s volcanic plumes. Post-*Galileo* missions (e.g., *Cassini, New Horizons*, and *Juno*) have all broadened our understanding of Io, but a dedicated mission is needed to provide an optimized suite of instruments, encounters, and measurements to answer the most fundamental questions about Io’s interior, surface, atmosphere, and role within the evolution of Galilean satellites and broader the Jupiter System.

**Acknowledgments**

We gratefully acknowledge support from the University of Arizona Space Institute (UASI), and thank the USGS internal reviewers for their careful review of this manuscript. Part of this work was performed at the Jet Propulsion Laboratory (JPL), California Institute of Technology, under contract to the National Aeronautics and Space Administration (NASA; 80NM0018D0004). AGD acknowledges the support of the NASA New Frontiers Data Analysis Program under award 80NM0018F0612.

**Tables**

TABLE 1.

**Table 1.** Overview of the Science Objectives for the *Io Volcano Observer* (*IVO*) mission concept. Key instruments are described in Table 2 and Section 3.2.

| Science Questions | Measurement Objective | Key Instrument(s) | Data Products | How Data Products Address Each Science Objective |
|---|---|---|---|---|
| **A:** How and where tidal heat is generated inside Io | **A1a**. Tidal $k_2$ | Gravity Science (GS) | Deep Space Network (DSN) Doppler tracking data | Inversion determines tidal $k_2$ to ±0.1. $k_2$ ~0.1 indicates a solid Io, while a $k_2$~0.5 indicates a magma ocean (Bierson & Nimmo 2016; Park et al. 2011). |
| | **A1b**. Libration amplitude | NAC | Match and tie points collected on framing images from I1, I2, and other orbits | Analysis after Thomas et al. (2016). Large amplitude if there is a magma ocean ***and*** the lithosphere is rigid (Figure 3; de Kleer et al. 2019a; Van Hoolst et al. 2020). |
| | **A1c**. Magnetic induction | DMAG, PIMS, lab experiments | Vector samples and time-ordered PIMS data; lab results in format for National Data Center. | Determine if Io's mantle contains abundant magma, resolving controversy by constraining plasma effects (Khurana et al. 2011; Blöcker et al. 2018; Šebek et al. 2019). Lab experiments for effect of sulfur volatiles and melt distribution on electrical conductivity (Pommier et al. 2008). |
| | **A1d**. Lava composition | TMAP | Map-projected multispectral thermal images | Christiansen feature wavelength (Figure 7) or shoulder proportional to $SiO_2$ content (Greenhagen et al. 2010; Maturilli & Helbert, 2014). |
| | | NAC | Lava temperatures | Lava temperature constrains composition (McEwen et al. 1998; Davies et al. 2001; Keszthelyi et al. 2007; de Kleer et al. 2014) |
| **B:** How tidal heat is transported to the surface of Io | **B1a**. Lithospheric thickness + rigidity | GS, DMAG, PIMS, NAC | Same as ***A1a–A1c*** | A combination of three measurements tightly constrains these values (Figure 4) and enables a key test of the heat-pipe model (Moore 2001). |
| | **B1b**. Global and topographic mapping | NAC | Global mosaic at 500 m/pixel; Digital Terrain Models (DTMs) combining limb profiles and stereo-photogrammetry | Tectonic mapping (Ahern et al. 2017); flexure models; elastic/viscous response depends on heat flux (White et al. 2014; Nimmo et al. 2011; Steinke et al. 2020). |
| | **B2a**. Endogenic heat flow | TMAP | Multi-wavelength mosaics and derived thermal inertia and heat flow maps | Heat flow may vary systematically with location, depending on heat generation and transport (Figure 3) (Hamilton et al. 2013; Tyler et al. 2015). |
| | **B2b**. Active eruption parameters | NAC, TMAP | Map-projected mosaics and lava temperature maps; movies | Style and vigor of eruptions distinguish between magma ascent mechanisms (Davies 2007; Davies et al. 2010; |
| | | | | Keszthelyi et al. 2001, 2007; Spencer et al. 2020). |
| **C:** How Io is evolving | **C1a**. Rate of change of Io's orbit | GS | DSN radiometric ranging and Doppler tracking | Refine da/dt of Io, Europa, and Ganymede (de Kleer et al. 2019; Lainey *et al.* 2009; Dirkx et al. 2017). |
| | **C2a**. Neutral species near Io | INMS | Mass spectra | Composition and abundances of neutrals are key inputs to determining volatile budget; use to test atmospheric models. |
| | **C2b**. Monitor plumes | NAC | Co-registered full-disk images in clear and color bandpasses | Track plume locations, sizes, and temporal variability (McEwen & Soderblom 1983; Geissler & McMillan 2008). |
| | **C2c**. Eclipse gas emissions | NAC | Co-registered full-disk images and movies in clear and color bandpasses | Monitor gas emissions by location and time to constrain models for atmospheric properties and magnetospheric interactions (Moore et al. 2009; Geissler et al. 2001; de Kleer et al. 2019a). |
| | **C2d**. Plasma and magnetic fields | PIMS, DMAG | Time-ordered PIMS data and DMAG vectors | Constrain models for flow of mass and energy (Thomas et al. 2004; Bagenal & Delamere 2011). Analysis follows that planned for PIMS on *Europa Clipper*. |

TABLE 2.

**Table 2.** *IVO* and *IVO-NF* Baseline mission experiments (FOV = Field Of View; TBD = To Be Determined).

| Baseline Mission or SCI | Experiment | Specifications | Science Drivers |
|---|---|---|---|
| *IVO/IVO-NF* | Gravity Science (GS) | 2-way Doppler tracking | Measure tidal deformation and Io's orbital evolution |
| *IVO/IVO-NF* | Narrow-Angle Camera (NAC) | 1.2° × 2.3° FOV, 10 µrad/pixel, 2048 × 4096 pixels, color stripes for pushbroom imaging in 12 bands; framing images | Measure libration, map Io, measure lava temperatures, monitor activity, topography |
| *IVO/IVO-NF* | Thermal Mapper (TMAP) | 7.2° × 5.1° FOV, 125 mrad/pixel, 1024 × 768 microbolometer array, 8 spectral bandpass stripes 4.5–14 µm; radiometer | Measure heat flow, lava eruption models, composition of silicates |
| *IVO/IVO-NF* | Dual Fluxgate Magnetometer (DMAG) | Low-noise sensors with high range and sensitivity on 3.5 m boom | Magnetic induction from subsurface magma and magnetospheric interactions |
| *IVO/IVO-NF* | Plasma Instrument for Magnetic Sounding (PIMS) | 2° × 90° conical FOV, 0.05–2.0 keV (electron), 0.05–6.0 keV (ions), | Measure plasma variations to interpret magnetic induction and plasma interactions |
| *IVO/IVO-NF* | Ion and Neutral Mass Spectrometer (INMS) | Mass range 1–1000 amu/q with M/ΔM = 1100 | First comprehensive measurements of ion and neutral species escaping from Io |
| *IVO* Student Collaboration Instrument | Student Wide Angle Camera (SWAC) | 50° × 25° FOV panchromatic (450–1000 nm) framing camera with a 46.2 mm focal length and an F/6 refractor | Stereo mapping, limb profiles, context images |
| *IVO-NF* | Wide Angle Camera (WAC) | 48° × 24° FOV, 218 µrad/pixel, 2048 × 4096 pixels, color stripes for pushbroom imaging in ~7 (TBD) bands; framing images | Stereo and color mapping, limb profiles, context images, lava temperature constraints |
| *IVO-NF* | Surface Dust Analyzer (SUDA) | Dust mass spectrometer, M/ΔM 150–300, mass range 1–150 amu | Composition and physical properties of dust particles near Io |

TABLE 3.

**Table 3.** Derived products generated from Discovery-class *IVO* instruments to meet Level-1 science requirements and provide foundational datasets. Products were planned to be delivered within six months after receiving raw and calibrated products or >6 months after the end of mission. "Foundational" refers to products that will have broad utility in the Planetary Sciences community (Williams et al. 2021).

| | | **Associated Science Goal(s)** | | | |
|---|---|---|---|---|---|
| **Instrument/Experiment** | **Derived Product** | **A** | **B** | **C** | **Foundational** |
| Gravity | Gravity model | + | + | | + |
| NAC | Polar geodetic networks | + | + | | + |
| NAC + SWAC | Global geodetic network | + | + | | + |
| NAC + SWAC | Smithed pointing | + | + | + | + |
| NAC + SWAC | Libration parameters | + | + | | + |
| TMAP | Global emissivity parameter maps | + | + | | |
| NAC | Local color temperature maps | + | | | |
| NAC + SWAC | Limb profiles | | + | | + |
| NAC + SWAC | Local and regional stereo topographic models | | + | | + |
| NAC | Local panchromatic lava movies | + | + | | |
| NAC + SWAC | Global shape model | + | + | + | + |
| NAC + SWAC | Catalog of geologic features | + | + | + | |
| NAC + SWAC + TMAP | Nomenclature | + | + | + | + |
| TMAP | Local 3-temperature maps | | + | | |
| TMAP + NAC | Global thermal inertia and albedo maps | | + | | |
| TMAP + NAC | Hot spot catalog | | + | + | |
| NAC | Full-disk and local multispectral mosaics | + | + | + | |
| NAC | Polar multispectral controlled and orthorectified mosaics | + | + | + | |
| NAC | Global panchromatic orthorectified mosaic | + | + | + | + |
| Spacecraft radio science | Updated orbital parameters | | | + | + |
| INMS | Ion and neutral density profiles | | | + | |
| NAC + TMAP | Multispectral global maps of Io | | + | + | |
| NAC | Multispectral eclipse movies | | + | + | |

**Science Goals:**

Goal A—Determine how and where tidal heat is generated inside Io.
Goal B—Understand how tidal heat is transported to the surface of Io.
Goal C—Understand how Io is evolving

TABLE 4.

**Table 4.** Raw, calibrated, and derived products that could be created from the NAC and WAC data (modified from Turtle et al. in press).

| Data Product | Description |
|---|---|
| Raw data | Uncompressed images |
| Map-projected framing images | Radiometric and geometric calibrations applied, including jitter corrections if needed; map projected |
| Map-projected pushbroom image segments | Radiometric and geometric calibrations applied, including jitter corrections if needed; map projected |
| WAC and NAC Digital Terrain Models (DTMs) | DTMs from WAC 3-line stereo, planned NAC stereo (good convergence angles and matching illumination), plus orthoimages and slope maps |
| Global and regional panchromatic mosaics | Mosaics incrementally improved after each flyby, preliminary geometric control |
| Global and regional color mosaics | Mosaics in ≤6 colors and clear, preliminary geometric control, NAC and WAC color |
| Shape of Io | Limb fit solutions |
| NAC and WAC regional panchromatic mosaics | Mosaic data from multiple flybys to cover regions of interest at better resolution than global mosaic |
| Global mosaic for morphology | Global mosaic of best images with high incidence angles to accentuate topography |
| Stereo anaglyphs | Qualitative viewing of topography. |
| 3-band color products | RGB natural color; CLR/1MC, 756/889, GRN/590 band ratios for lava temperatures |
| Merged color | Merge color mosaics with higher-resolution panchromatic mosaics |

TABLE 5.

**Table 5.** Comparison of Discovery- and New Frontiers-class *IVO* mission concepts for Io (See Table 2 for definition of Science Experiment acronyms; TBD = To Be Determined).

| | *IVO* | *IVO-NF* |
|---|---|---|
| Number of Io encounters | 10 | 20 |
| Lowest encounter altitude | 200 km | 50 km |
| Jupiter tour duration | 3.5 years | 3.5 years |
| Total data volume | 200 Gb | >400 Gb |
| Downlink | X-band | X-band and Ka-band |
| Baseline Science Experiments | GS, NAC, TMAP, DMAG, PIMS | GS, NAC, TMAP, DMAG, PIMS, WAC, SUDA |
| Student Collaboration | SWAC | TBD |
| Technology Demonstration Option (TDO) | RUSHeS | TBD |
| Static gravity science | Minimal | Much better (degree-5) |
| Dust abundance and composition | None | SUDA |
| Atmospheric and plume composition, including isotope ratios | Limited to INMS at 200 km or higher | INMS and SUDA down to 50 km |
| Mapping and monitoring Io | Near-global coverage that would address Discovery-level STM Goals | Enhanced near-global coverage, with higher resolution imaging, more repeat flyby coverage, and lower flybys later in the tour that together meets more ambitious New-Frontiers STM Goals, relative to *IVO* |

**Figures**

FIGURE 1.

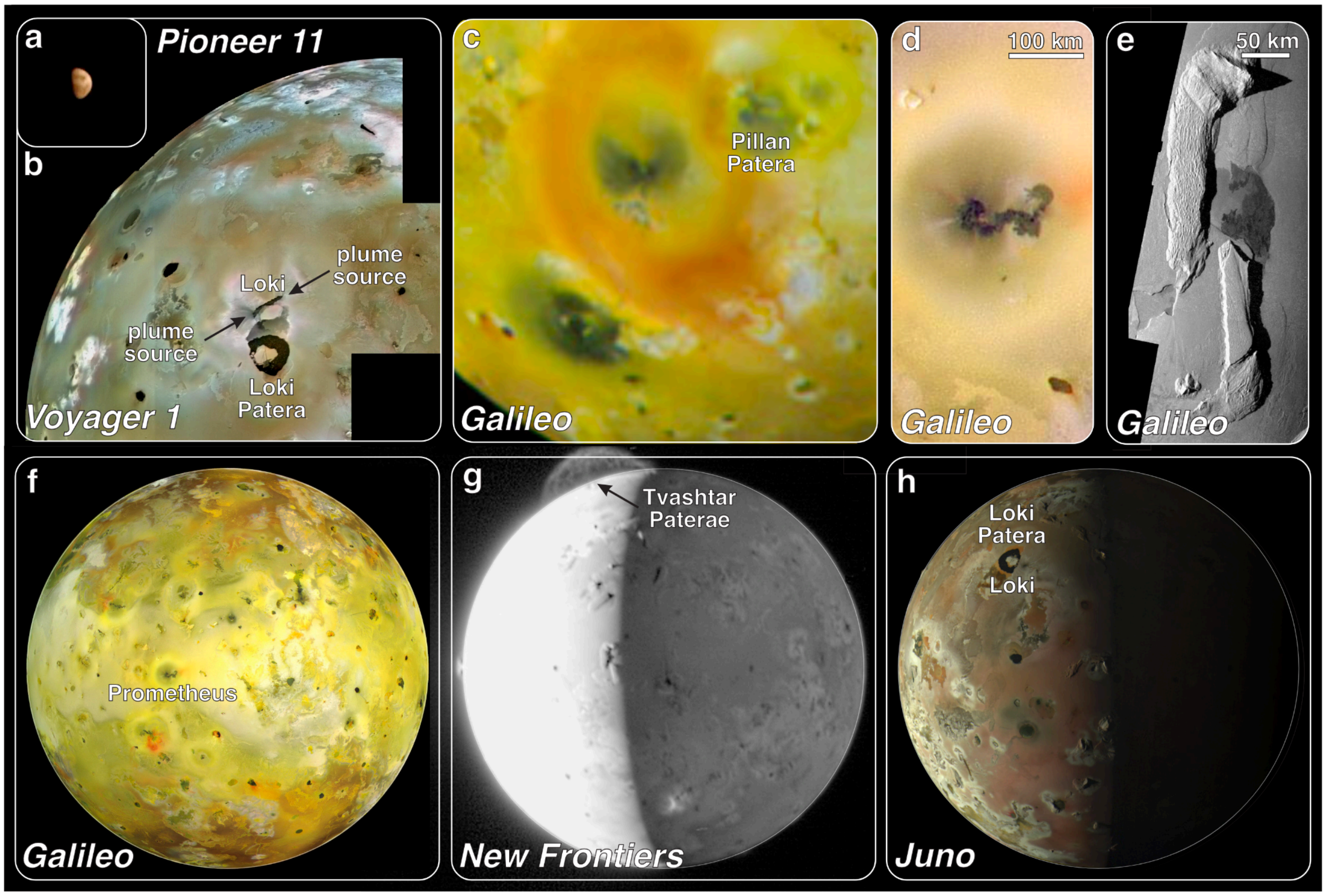

**Figure 1.** Views of Io. (a) *Pioneer 11* image of Io captured on 2 December 2 1974 from a distance of 756,000 km (Credit: NASA/JPL). (b) Loki and Loki Patera shown in a *Voyager 1* image mosaic (Credit: NASA/JPL/USGS). (c) Magnified view of part of a *Galileo* Color image of Io (11.8 km/pixel) captured 2 July 1999. This image shows Pele, a small lava lake that has produced a large, red ring that reaches as far as 600 km from the central vent, with overprinting by a recent dark tephra deposit from Pillan Patera (Credit: NASA/JPL/NOAO/Jason Perry). (d) *Galileo* view of Prometheus, which includes a large lava flow-field and secondary plumes associated with lava-sulfur frost interactions along the flow's margin (Credit: NASA/JPL). (e) *Galileo* view of Hiʹiaka Montes, the dark area near the center of this "pull apart" mountain is a wide volcanic crater called a patera. (Credit NASA/JPL). (f) *Galileo* acquired its highest resolution images of Jupiter's moon Io on 3 July 1999 during its closest pass to Io since orbit insertion in late 1995. This color

mosaic uses the near-infrared, green and violet filters to approximate“true color” (i.e., what the human eye would see). Most of Io’s surface has pastel colors, punctuated by black, brown, green, orange, and red units near the active volcanic centers. (g) *New Horizons* image of Io, showing a large volcanic plume erupting from Tvashtar (Image: PIA09248; Credit: NASA/APL). (h) Loki and Loki Patera shown in a *Juno* image captured during its 57$^{th}$ flyby of Jupiter, in December 2023 (Image: PJ57, Credit: NASA/JPL-Caltech/SwRI/MSSS/Kevin M. Gill). Note in panels (f) and (g), North is toward the top of the page, whereas the *Juno* image shown in (h) views looks toward the North Pole and the Northern Hemisphere.

FIGURE 2.

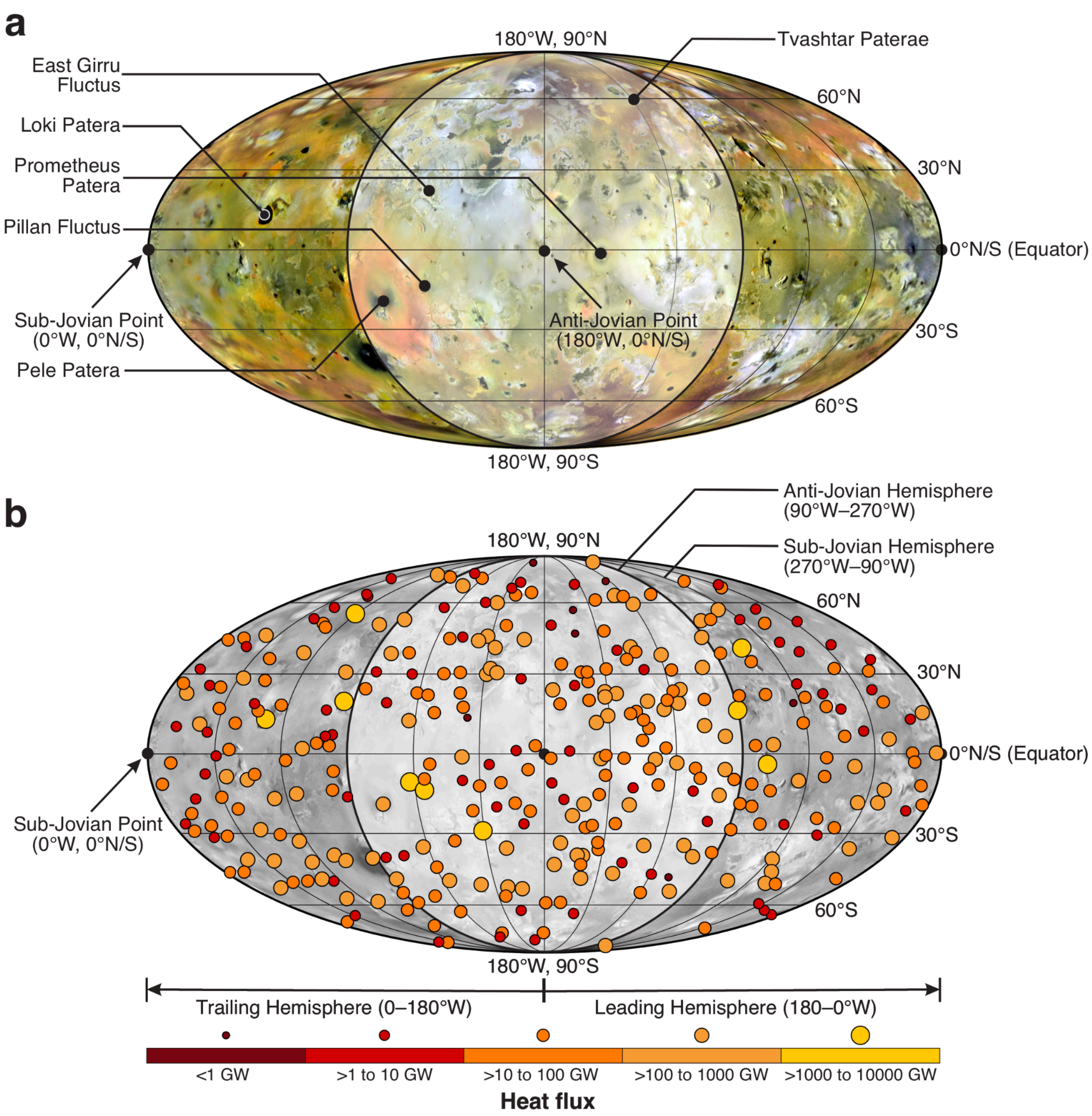

**Figure 2.** Sources of enhanced thermal emission on Io. Io is an extremely active volcanic world with over 343 active thermal sources, nearly all of which are more energetic than the most active eruptions on Earth today. For example, Kīlauea Volcano, on the island of Hawaiʻi, has a mean energy output <1 GW, which is less than nearly all volcanic systems on Io. (a) Global *Voyager-Galileo* global mosaic for Io, with the locations of key volcanic

systems discussed in the text. (b) Thermal emission (i.e., heat flux) for volcanic systems on Io (Davies et al. 2024b).

FIGURE 3.

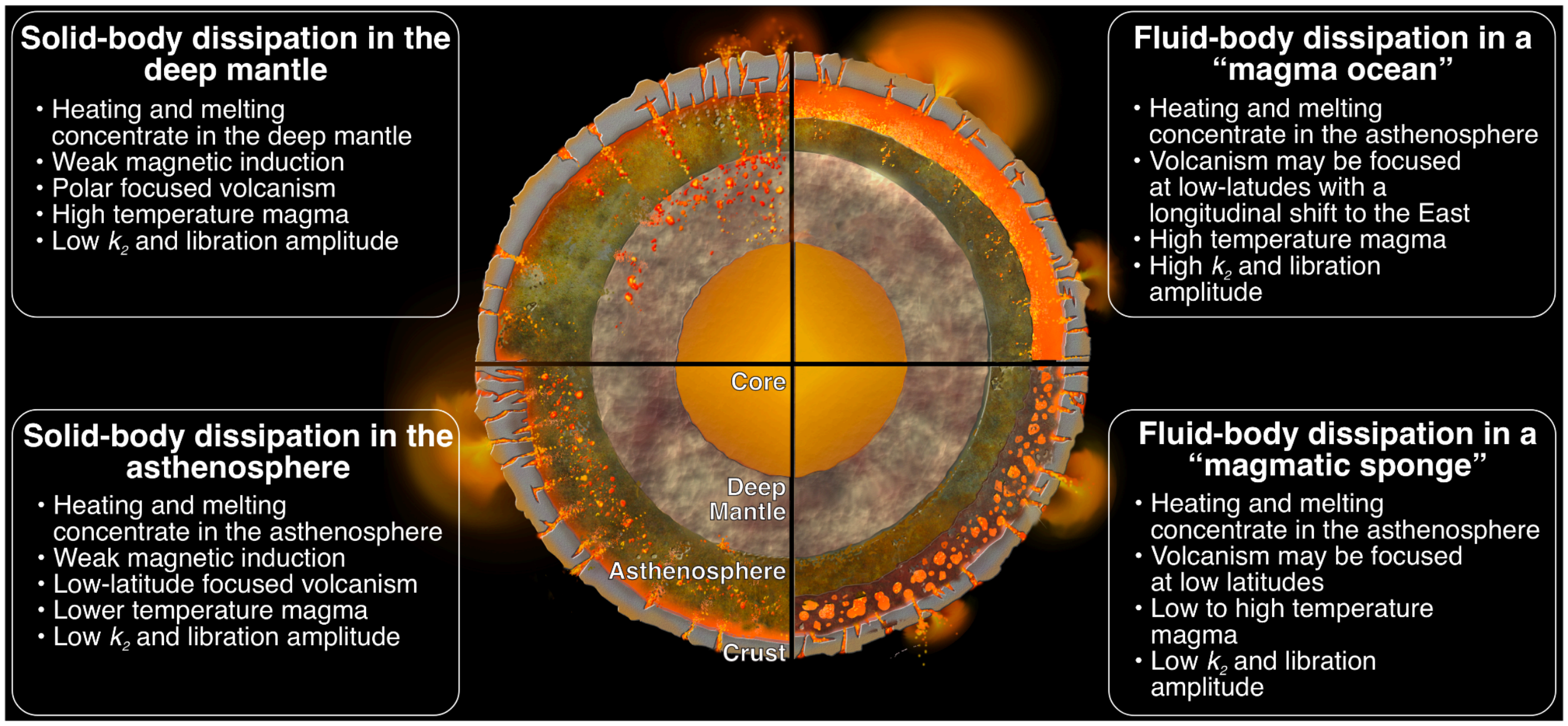

**Figure 3.** Conceptual models for the interior of Io, depending on where the bulk of tidal dissipation occurs (after de Kleer et al. 2019a; Breuer et al. 2022).

FIGURE 4.

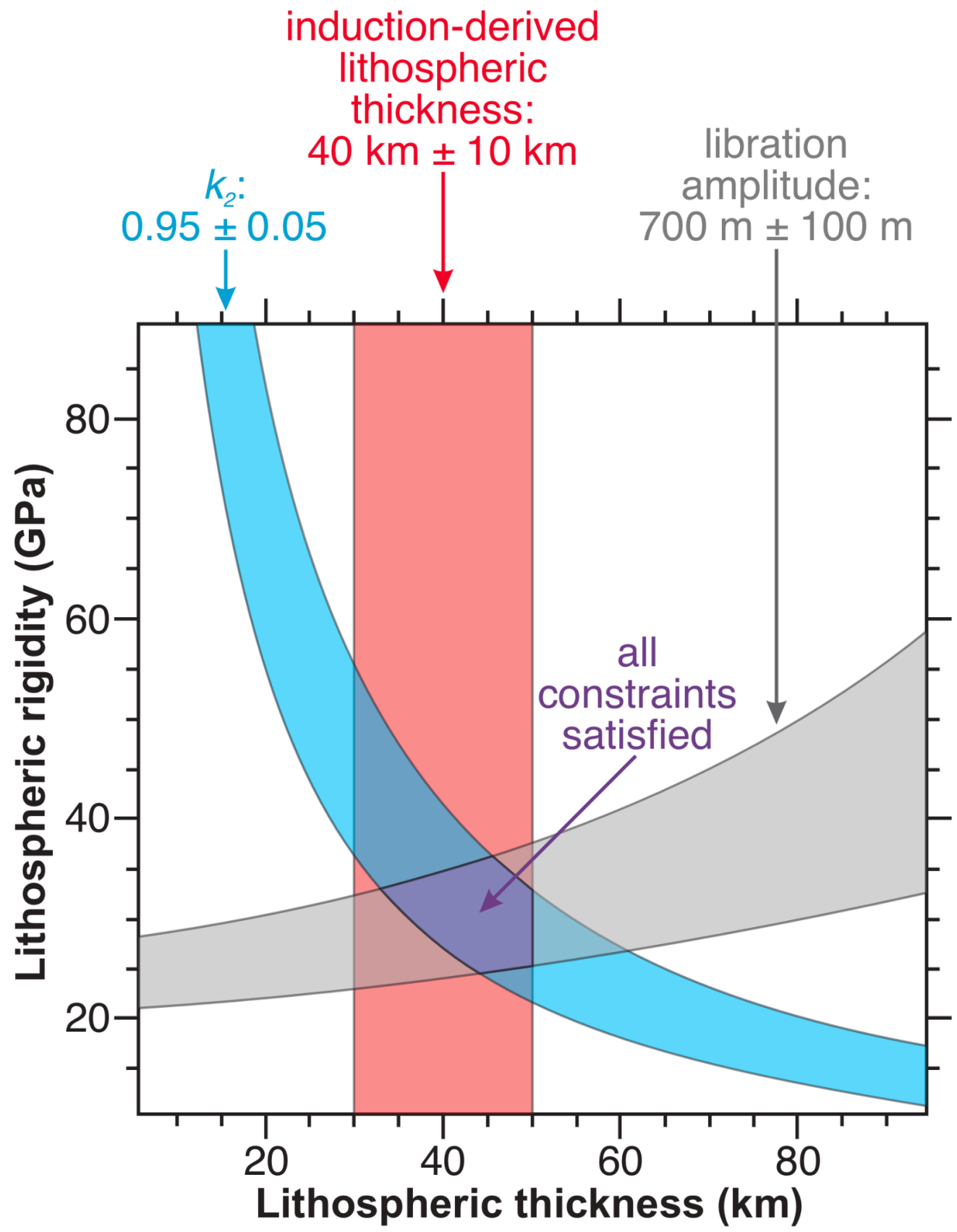

**Figure 4.** Constraints on Io's average lithospheric thickness and rigidity based on measurements of $k_2$, magnetic induction, and libration amplitude. *IVO* would precisely determine average lithospheric thickness and rigidity, which—combined with new magnetic induction measurements—would provide the constraints necessary to assess whether or not Io has an internal magma ocean (Keane et al. 2022). Use of multiple, independent, constraints is one of the most powerful techniques in geophysics.

FIGURE 5.

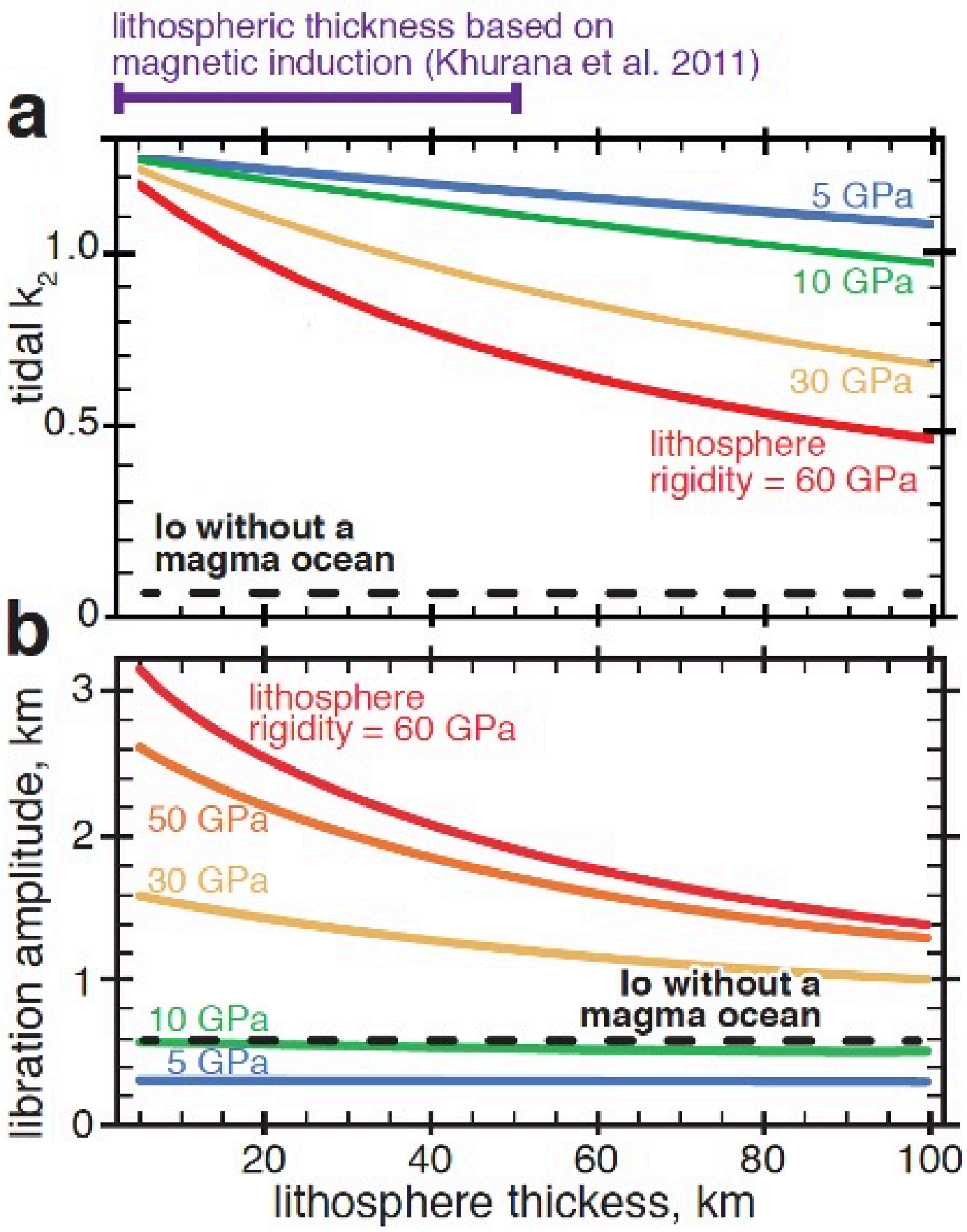

**Figure 5.** Measurement of (a) tidal $k_2$, (b) libration amplitude, and magnetic induction provide independent tests for a magma ocean (see Figure 3). Tidal $k_2$ and the peak-to-peak libration amplitude of Io are calculated assuming an elastic lithosphere overlying a magma ocean (Van Hoolst et al. 2020). Io's lithosphere is expected to be rigid (>30 GPa) and surprisingly thick (10–50 km) based on the volcanic heat-pipe model and to support mountains up to 17 km high. The peak-to-peak libration amplitude is two times the typically quoted value of displacement from the mean. Multifrequency magnetic sounding will measure the thickness of Io's lithosphere and constrain models for the thickness and conductivity of the magma-rich layer.

FIGURE 6.

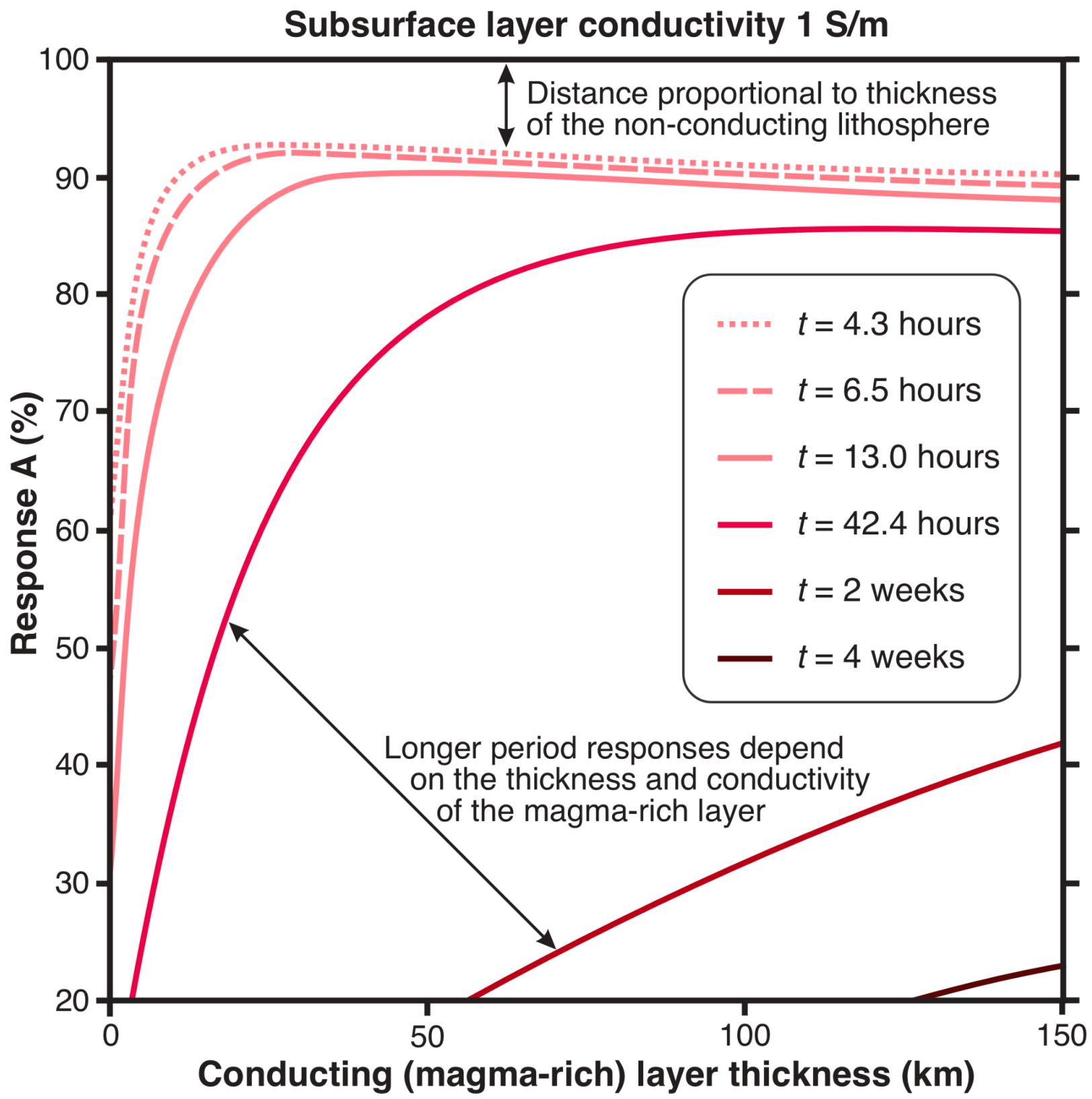

**Figure 6.** The normalized response of Io (i.e., ratio of the induced polar field at Io's surface to the strength of the inducing field) for a range of magma-rich layer thicknesses for six periods ranging from 4.3 hours to 4 weeks.

FIGURE 7.

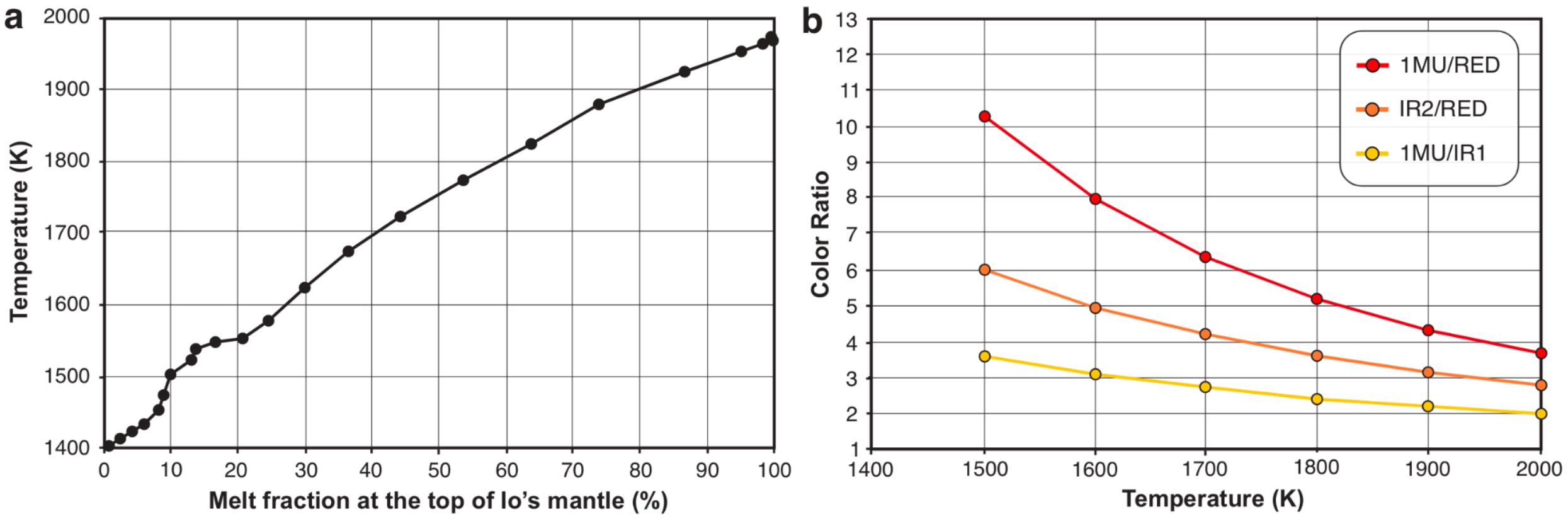

**Figure 7.** Melt fraction varies as a function of mantle temperature, so measuring the temperature of the lava as it erupts provides constraints on the state of Io's interior. (a) Plot of melt fraction at the base of the crust as a function of temperature assuming Io has a chondritic composition similar to other bodies in the solar system (Keszthelyi et al. 2007). Constraining the temperature of the erupting lava to ±50 K would place useful constraints on the state of Io's interior. However, lava cools rapidly so isolating the temperature of the lava as it exits the ground is very challenging. *IVO* would rely on two different strategies to meet this challenge. First, large lava fountains are regularly observed erupting on Io, providing a direct view of lava as it first erupts. Second, lava tubes are known to exist on Io, so skylights—locations where the roof of the tube has collapsed—should also exist. If these can be imaged, they provide a view of lava that has traveled away from the vent with minimal cooling. In both cases, in order to isolate the hottest subpixel region, color temperatures would be used. Color temperatures are obtained from ratios of the intensity of the incandescent glow as seen in different spectral bands. (b) Ratios of the red and near infrared bandpasses for the NAC provide reliable temperature estimates across the full range of plausible lava temperatures. The plotted bandpasses are 1MU: 950–1100 nm, IR2: 850–950 nm, IR1: 800–850 nm, RED: 650–800 nm. Using shorter wavelengths (e.g., green and orange) or narrow bandpasses allows identification of more exotic lava types that have been speculated to exist on Io (Kargel et al. 2003). The direct measurements of lava temperature would be tested with additional observations such as the silica content of the lava derived from thermal infrared spectroscopy (following the work of Greenhagen

et al. 2010) and the mix of volcanic gasses measured by the mass spectrometer (following the work of Spencer et al. 2000).

FIGURE 8.

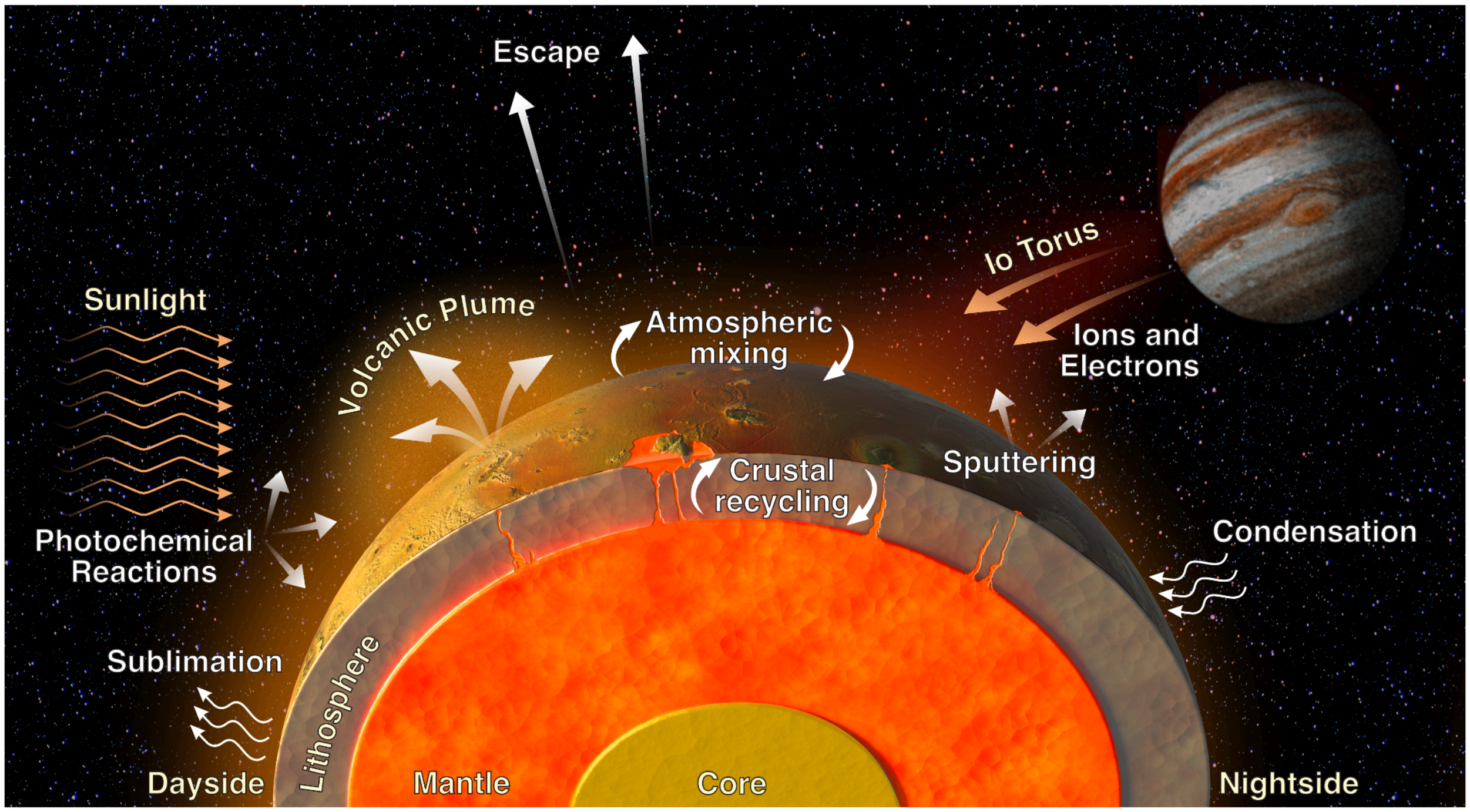

**Figure 8.** Volatile fluxes on Io, such as those illustrated here, are poorly understood. However, *IVO*'s INMS and other Objective C2 measurements (see Table 1) would provide important constraints for validating models used to determine rates for these processes as a necessary first step in understanding the long-term evolution of Io's volatile inventory and mass exchange within the Jovian System. In this figure, processes are labelled in white font.

FIGURE 9.

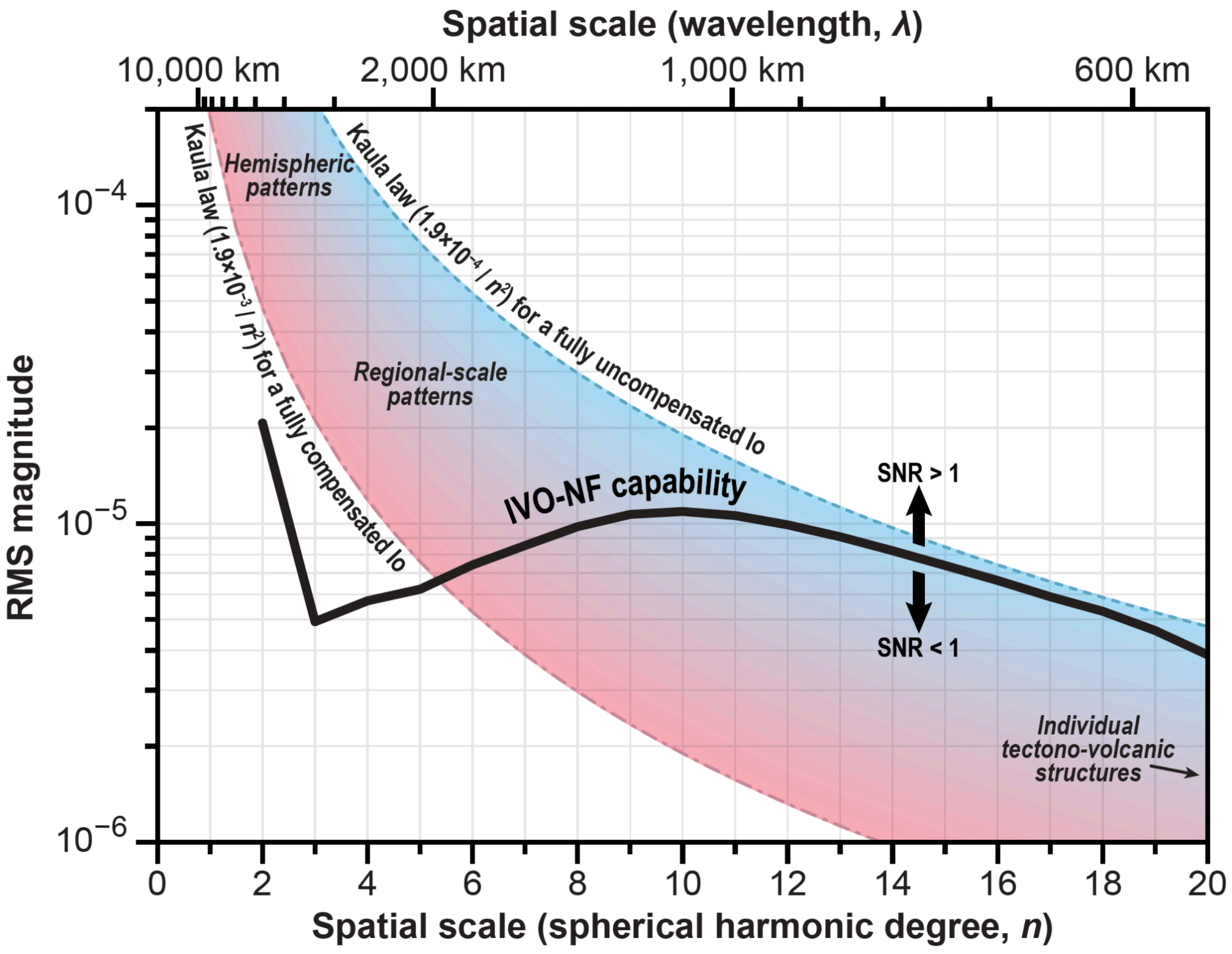

**Figure 9.** The expected quality of the global static gravity field for *IVO-NF*. The red/blue colored swath shows the range of plausible gravity fields of Io spanning the range of possible compensation states. The thick black line indicates the expected error spectrum from *IVO-NF*. Gravity signals above this line would be detectable with a signal-to-noise (SNR) greater than one. The present gravity field of Io is only resolved to degree-2. *IVO-NF* would enable detection of regional-scale gravity anomalies, and constrain the deeper interior structure of Io.

FIGURE 10.

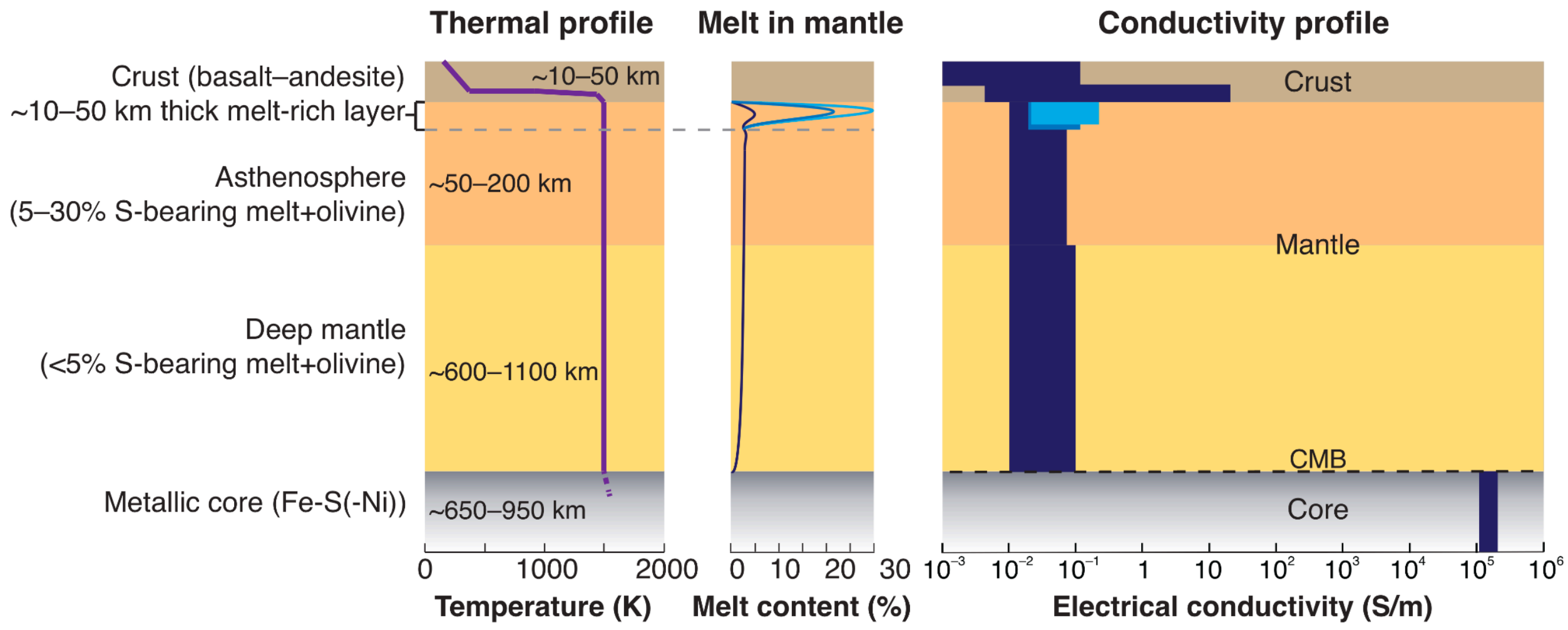

**Figure 10.** Electric conductivity of Io's interior, based on the thermal profile and composition and electrical laboratory studies. Left panel: composition and thermal profile across Io. Middle panel: Melt content and distribution in the silicate mantle, considering different melt contents (5%, 20%, 30%) in the top part of the asthenosphere. Right panel: corresponding conductivity-depth profile. In the hypothetical melt-rich layer in the asthenosphere, dark blue, intermediate blue, and light blue data correspond to 5%, 20% and 30% melt, respectively, with the width of the lines representing modeled electrical conductivities for each layer. Note that the high electrical conductivity values in the lower crust reflect the combination of the sharp increase in temperature at this depth and the composition of pure basalt, which has conductivity higher than olivine (i.e., mantle material) at 1500°C. CMB = Core-Mantle Boundary.

FIGURE 11.

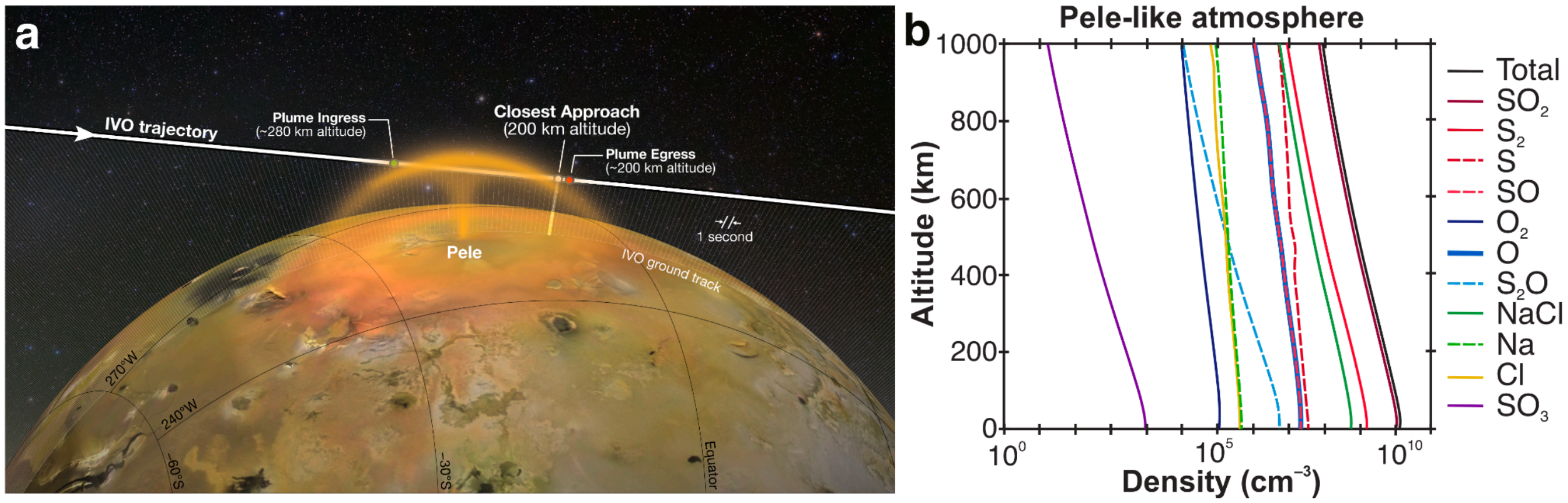

**Figure 11.** (a) Schematic of *IVO-NF*'s trajectory through a plume erupted from Pele Patera, with (b) expected atmospheric profile above the volcano (Moses et al. 2002).

FIGURE 12.

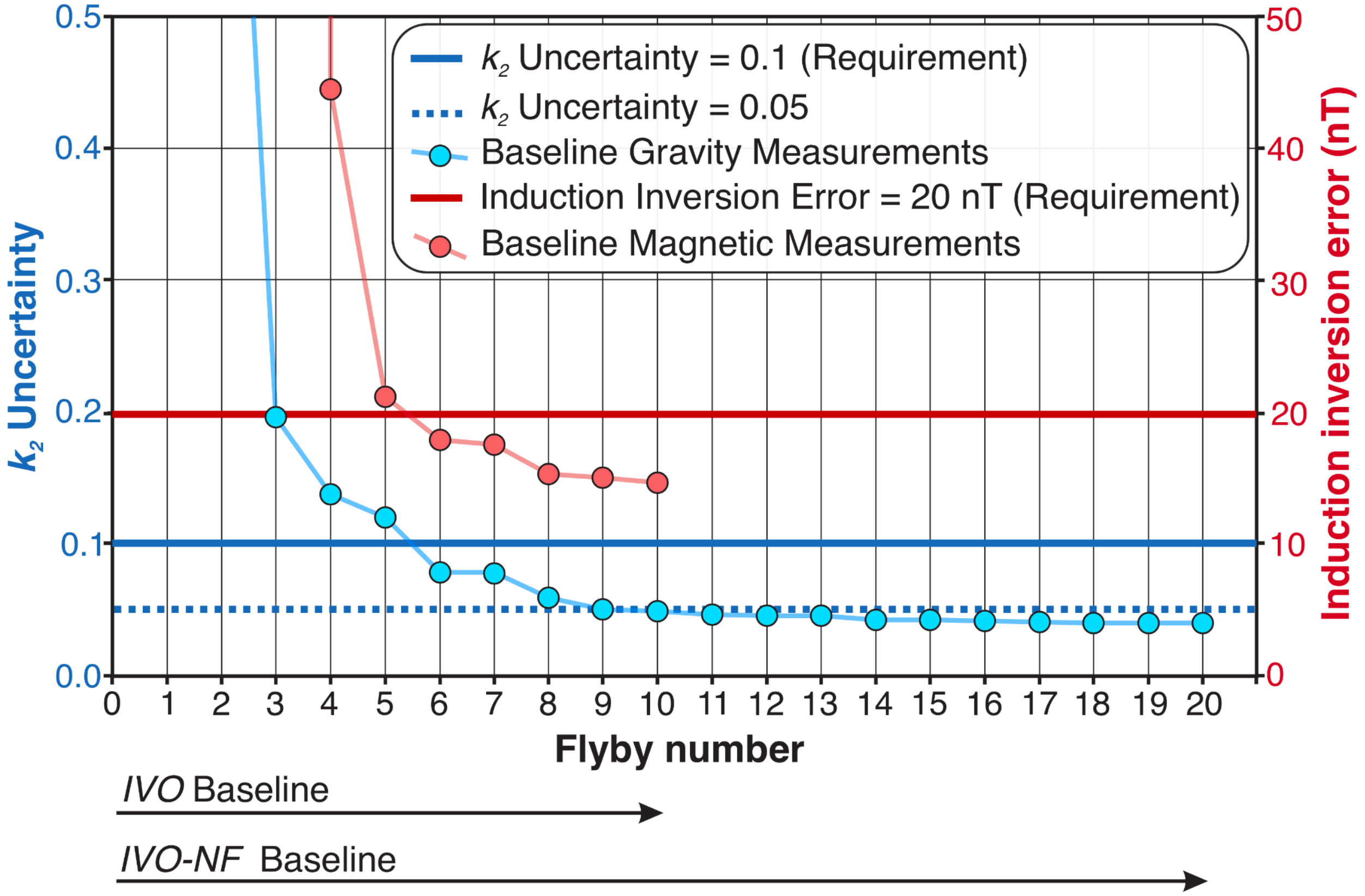

**Figure 12.** Improvement in gravity science ($k_2$ uncertainty) and magnetic (induction inversion error) as a function of flyby number. The *IVO* Baseline mission has 10 flybys of Io, whereas the *IVO-NF* Baseline mission has 20 flybys, using the *IVO-NF* Iteration-5 tour. Estimates of induction inversion error are unique for each tour and have yet to be completed for *IVO-NF*, but are expected to be substantially better than the *IVO* Baseline over the course of the 20 flyby tour.

FIGURE 13.

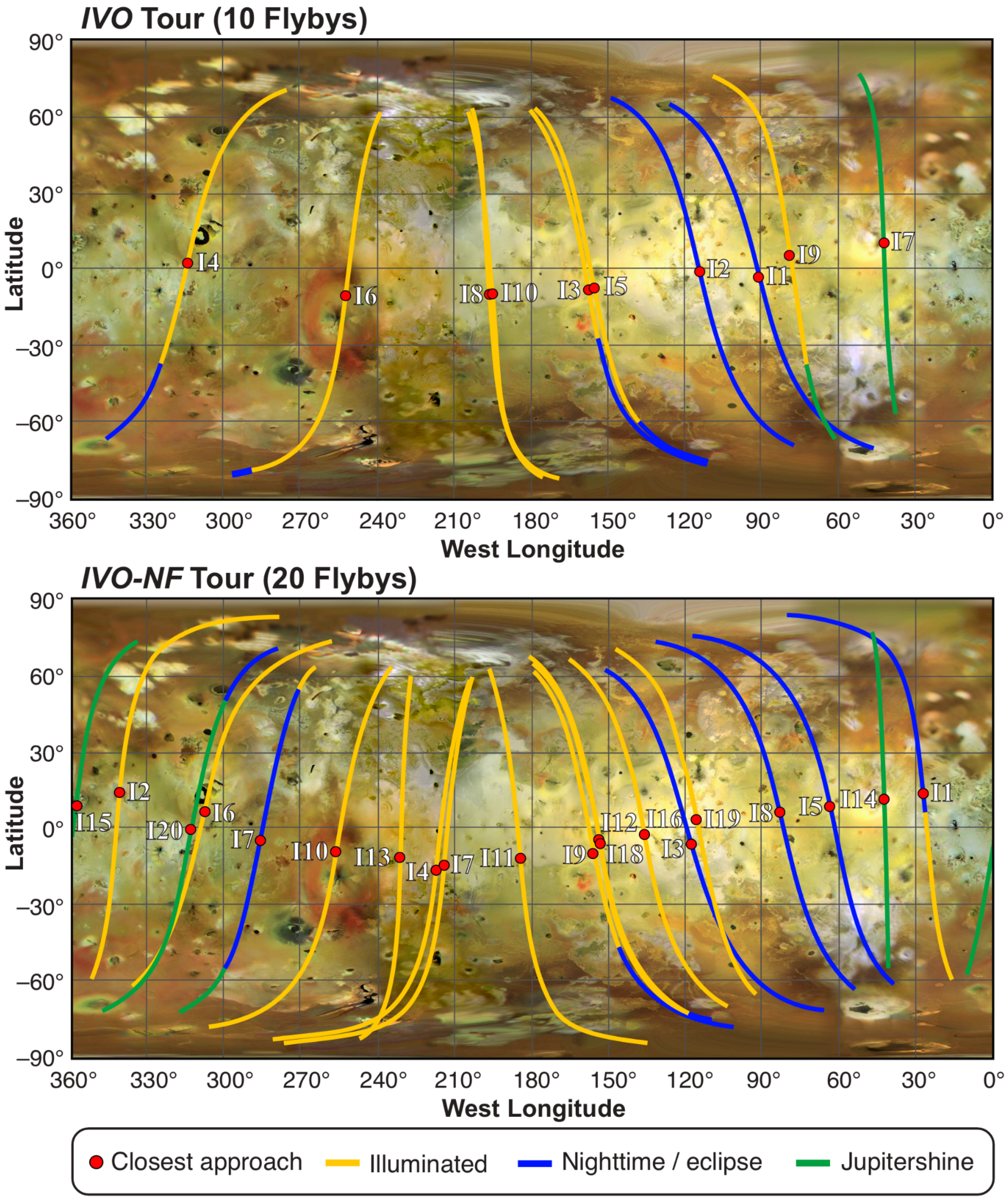

**Figure 13.** Top: *IVO* Baseline tour. Bottom: *IVO-NF* Baseline tour example (Iteration-5). Trajectories are plotted over the *Galileo–Voyager* image mosaics

produced at a spatial resolution of 1 km/pixel (Williams et al. 2011), with flybys labeled I1–I10 for *IVO* and I1–I20 for *IVO-NF*.